\documentclass[10pt,twocolumn,english,prd,superscriptaddress,nofootinbib,preprintnumbers,showpacs,floatfix]{revtex4-2}
\usepackage{amsmath,amssymb,amsthm,hyperref}
\usepackage[utf8]{inputenc}
\usepackage{graphicx}
\usepackage{xcolor}

\begin{document}
	

\title{Radiation reaction in the classical relativistic St{\o}rmer problem}

\author{Francisco S. N. Lobo}
\email{fslobo@ciencias.ulisboa.pt}
\affiliation{Instituto de Astrof\'{i}sica e Ci\^{e}ncias do Espa\c{c}o, Faculdade de Ci\^{e}ncias da Universidade de Lisboa, Edificio C8, Campo Grande, P-1749-016 Lisbon, Portugal }
\affiliation{Departamento de F\'{i}sica, Faculdade de Ci\^{e}ncias da Universidade de Lisboa, Edif\'{i}cio C8, Campo Grande, P-1749-016 Lisbon, Portugal}

\author{Tiberiu Harko}
\email{tiberiu.harko@aira.astro.ro}
\affiliation{Department of Physics, Babe\c s-Bolyai University, Kog\u alniceanu Street,
	Cluj-Napoca 400084, Romania,} 
\affiliation{Astronomical Observatory, 19
	Cire\c silor Street, 400487 Cluj-Napoca, Romania}
	
\date{\today}

\begin{abstract}
	We extend the classical relativistic St{\o}rmer problem by incorporating
	radiation reaction through the Landau--Lifshitz formulation, thereby providing
	a self-consistent description of dissipative charged-particle motion in a
	static dipole magnetic field. We derive the complete dimensionless equations
	and distinguish them from a reduced drag-only model that preserves the exact
	energy-loss law while omitting directional effects associated with magnetic
	field gradients. For planar motion, the reduced system yields exact
	instantaneous evolution laws for the particle energy and canonical angular
	momentum, together with averaged transport equations for regular bound
	librations. When the motion is constrained to the instantaneous circular
	branch, the secular evolution can be integrated in closed form and approaches
	a simple large-radius power law. This analytical solution provides a useful
	benchmark, although the circular branch is radially unstable and therefore
	does not describe generic nearby trajectories. Numerical integrations further
	illustrate the nonuniform dissipative deformation of planar rosette-like
	orbits. In three dimensions, the complete Landau--Lifshitz force produces
	local exponential damping of small vertical perturbations, with the leading
	contribution arising from the field-gradient term absent from the reduced
	model. These exact, averaged, conditional, and local results establish a
	controlled analytical framework for studying radiation-driven phase-space
	transport in strongly inhomogeneous magnetic fields and provide a foundation
	for future global simulations, kinetic descriptions, and calculations of the
	associated electromagnetic emission.
\end{abstract}


\maketitle
\tableofcontents

\section{Introduction}
\label{sec:intro}

The St{\o}rmer problem was originally developed to describe the motion of
charged particles in the Earth's dipole magnetic field and has played a
foundational role in the theory of auroral phenomena, geomagnetic trapping,
and radiation belts
\cite{St1,St2,St3,St4,St4a,St5,St6,St7}. Its classical and relativistic
forms have since been studied from analytical, numerical, and observational
perspectives
\cite{B1,B2,B3,Int,Dilao,VA1,VA2,Schust,How,Dull,In,In1,Epp0,Epp,Hal,
	Mark,Ozturk,Pina,Kol,Leg,Ersh,Asadi,Ersh1,Moc,Pap,Bur,Bur1}.
In the classical relativistic St{\o}rmer problem (CRSP), a charged particle
moves under the Lorentz force of a prescribed static magnetic dipole. Because
the magnetic field performs no work, the Lorentz factor is conserved, while
axial symmetry provides a conserved canonical angular momentum
\cite{Jack,LL}.

A recent analysis of the conservative CRSP obtained exact parametric
solutions for planar motion and complemented them with a systematic numerical
study of three-dimensional trajectories and of the radiation emitted by the
accelerated charge \cite{Harko:2026tev}. Those results clarified important
features of the relativistic dipole problem, but the particle trajectories
were still determined by the Lorentz force alone. The recoil associated with
the emitted radiation was not included in the orbital dynamics.

Radiation reaction becomes relevant whenever the energy radiated by an
accelerated charge accumulates sufficiently to modify its motion. This may
occur in strongly magnetised or highly relativistic environments, where the
radiative timescale can compete with the timescale of particle confinement or
transport. A self-consistent treatment must then distinguish the radiation
calculated from a prescribed trajectory from the back-reaction of that
radiation on the trajectory itself.
The astrophysical connection between relativistic charged particles, stellar dipole magnetic fields, and synchrotron emission has a long history, including the early analysis by Thorne of synchrotron radiation from stars endowed with strong dipole magnetic fields \cite{Th}.

The covariant point-particle description is provided by the
Lorentz--Abraham--Dirac equation, whose third-order character permits runaway
and pre-accelerating solutions. Landau and Lifshitz instead introduced a
reduction-of-order procedure in which the acceleration entering the
self-force is replaced by its Lorentz-force value. The resulting
Landau--Lifshitz (LL) equation is second order and is accurate to first order
in the classical radiation-reaction time, within the regime in which the
self-force remains a perturbative correction \cite{LL,DiPiazza:2011tq}.

Radiative effects in compact-object magnetospheres have been explored through
analytical approximations, numerical orbit integrations, and kinetic
simulations. Barkov and Lyutikov studied relativistic particles trapped in the dipolar magnetospheres of pulsars and magnetic white dwarfs, incorporating synchrotron radiative damping together with adiabatic magnetic-mirror forces. They identified bouncing, precipitating, and freezing trajectories and calculated their associated multi-frequency emission patterns \cite{Barkov:2025uag}.
P\'{e}tri developed an analytical particle pusher based on the reduced LL
equation and applied it to rotating dipole fields, finding that radiation
reaction affects electrons much more strongly than heavier particles and can
substantially reduce their attainable Lorentz factors
\cite{Jerome:2022emr}. Tomczak and P\'{e}tri extended this approach to
ultra-strong neutron-star fields and found radiation-limited electron Lorentz
factors of order $10^{10.5}$ in their rotating-vacuum-dipole models
\cite{Tomczak:2023ftp}. In a relativistic dipole magnetosphere embedded in
Schwarzschild spacetime, Stuchl\'{i}k, Vrba, Kolo\v{s}, and Tursunov showed
numerically that the outcome depends on the sign of the electromagnetic
coupling and on the initial orbital latitude: some repulsive configurations
widen and approach the equatorial plane, whereas others, as well as the
attractive configurations studied there, evolve towards the stellar surface
\cite{Stuchlik:2024tlu}. Particle-in-cell studies have likewise demonstrated
the close connection between magnetospheric structure, particle acceleration,
and high-energy emission
\cite{Philippov:2013tpa,Cerutti:2015hvk}.

These studies establish the physical importance of radiation reaction, but
they do not provide an analytical account of LL-driven transport in the
static, purely magnetic CRSP. In particular, it remains useful to determine
which properties follow exactly from the complete LL force, which require a
reduced drag-only model, and which are valid only near special families of
conservative trajectories. This distinction is especially important in a
dipole field, where strong spatial inhomogeneity makes the field-gradient
part of the LL force potentially relevant to the direction of the momentum,
even though it performs no work.

The purpose of this paper is to develop such a controlled formulation. We
derive the complete dimensionless LL equations for a charged particle in a
static magnetic dipole and identify the dimensionless parameter that measures
the strength of radiation reaction relative to the conservative St{\o}rmer
dynamics. We adopt the sign convention $eM>0$, which fixes the orientation of
the electromagnetic coupling, and distinguish throughout between the
complete LL equation and a reduced model that retains only the terms
quadratic in the magnetic field.

For equatorial motion, the reduced system yields exact instantaneous
evolution laws for the Lorentz factor and the canonical angular momentum.
When the dissipative change over one conservative radial period is small,
these laws can be averaged to obtain a coupled secular transport system for
regular librational trajectories away from separatrices. The canonical
radial action provides a useful phase-space diagnostic, but it is not assumed
to be conserved by the dissipative flow.

A closed-form solution follows when the particle is constrained to the
instantaneous circular branch. Along this branch, radiative energy loss is
accompanied by an increase of the radius, and the resulting first-order
evolution can be integrated exactly. Its large-radius, nonrelativistic limit
exhibits a simple power-law behaviour. This solution is conditional rather
than generic because the conservative circular branch is radially unstable.
Numerical integrations of the reduced planar equations are therefore used to
illustrate the deformation of more general trajectories without assuming
universal inward or outward migration.

For three-dimensional motion, we return to the complete LL force. Linearising
about the instantaneous circular branch shows that small vertical
perturbations are locally damped and that the leading damping term originates
from the LL field-gradient contribution omitted from the reduced planar
model. A local action-angle description extends this result to weakly
nonlinear vertical oscillations under a frozen-background approximation. The
analysis establishes local transverse damping near the circular branch, but
not a global equatorial attractor for arbitrary three-dimensional motion.

The scope of the paper is therefore deliberately hierarchical. We obtain
exact dissipation identities for the complete LL dynamics, a reduced planar
model with an explicit domain of applicability, averaged transport equations
for regular planar librations, an exact conditional solution along the
circular branch, and a local transverse-stability result in three dimensions.
This organisation separates robust conclusions from modelling assumptions
and provides analytical benchmarks for future global numerical studies.

The paper is organised as follows. In Sec.~\ref{sec:model}, we present the
conservative CRSP, introduce the complete LL force, and derive the complete
and reduced dimensionless systems. Section~\ref{sec:planar} studies
equatorial motion, including the conditional circular-branch evolution, the
role of the field-gradient term, the exact analytical integration, averaged
phase-space transport, and representative numerical trajectories.
Section~\ref{sec:3D} analyses local transverse stability and the corresponding
vertical action-angle description, and clarifies the limits of the resulting
three-dimensional conclusions. We discuss the principal results and future
directions in Sec.~\ref{sec:discussion}. Throughout, we use Gaussian units
and retain explicit factors of $c$.

\section{Model and equations of motion}
\label{sec:model}

In the present Section we introduce the theoretical foundations of the relativistic St{\o}rmer problem, as well as the Landau-Lifshitz formulation of the radiation reaction force. Then the basic evolution equations of the St{\o}ormer problem with radiation reaction are written down, and formulated in a dimensionless form.   

\subsection{Conservative relativistic St\"{o}rmer problem}

We briefly recall the classical relativistic St{\o}rmer problem (CRSP),
which provides the conservative background for the dissipative analysis
developed below. The system describes a particle of mass $m$ and charge $e$
moving in the field of an ideal magnetic dipole
$\mathbf{M}=M\mathbf{e}_{z}$. A convenient vector potential is
\begin{equation}
	\mathbf{A}
	=
	\frac{M}{r^{3}}(-y,x,0),
	\qquad
	r=\sqrt{x^{2}+y^{2}+z^{2}},
	\label{eq:A_dipole}
\end{equation}
which generates
\begin{equation}
	\mathbf{B}
	=
	\boldsymbol{\nabla}\times\mathbf{A}
	=
	\frac{3Mz}{r^{5}}(x,y,z)
	-
	\frac{M}{r^{3}}\mathbf{e}_{z}.
	\label{eq:B_dipole}
\end{equation}
The field is stationary, axially symmetric, and strongly inhomogeneous.
Although the magnetic force changes only the direction of the particle
momentum, the spatial variation of the dipole field produces a nontrivial
phase-space structure containing bounded, escaping, regular, and chaotic
trajectories, depending on the initial data
\cite{Harko:2026tev}.

The relativistic equation of motion is
\begin{equation}
	\frac{d}{dt}\left(\gamma m\mathbf{v}\right)
	=
	\frac{e}{c}\,\mathbf{v}\times\mathbf{B},
	\qquad
	\gamma
	=
	\frac{1}{\sqrt{1-v^{2}/c^{2}}}.
	\label{eq:Lorentz}
\end{equation}
Since
$\mathbf{v}\cdot(\mathbf{v}\times\mathbf{B})=0$, the magnetic field
performs no work. Hence the particle energy
$E=\gamma mc^{2}$, and therefore $\gamma$, are constant. Axial symmetry
also implies conservation of the canonical angular momentum
\begin{equation}
	L_{z}
	=
	x\Pi_{y}-y\Pi_{x},
	\qquad
	\boldsymbol{\Pi}
	=
	\gamma m\mathbf{v}
	+
	\frac{e}{c}\mathbf{A}.
	\label{eq:Lz_dimensional}
\end{equation}
The distinction between mechanical and canonical angular momentum is
important: the latter contains the contribution of the vector potential
and is the quantity conserved by axial symmetry.

We introduce a reference length $R_{0}$ and the dimensionless variables
\begin{align}
	x&=R_{0}X,
	&
	y&=R_{0}Y,
	&
	z&=R_{0}Z,
	\nonumber\\
	\tau
	&=
	\frac{eM}{mcR_{0}^{3}}\,t.
	\label{eq:dimensionless_variables}
\end{align}
Here $\tau$ is a rescaled coordinate time, not a proper time.
We adopt the sign convention $eM>0$, so that increasing $\tau$ corresponds
to increasing physical time. If both signs of $eM$ are to be treated
simultaneously, one may instead define $\tau$ using $\lvert eM\rvert$ and
retain the factor $\operatorname{sgn}(eM)$ explicitly in the dimensionless
Lorentz-force terms.
Defining
\begin{equation}
	\mathbf{R}=(X,Y,Z),
	\qquad
	R=\sqrt{X^{2}+Y^{2}+Z^{2}},
\end{equation}
and
\begin{equation}
	\mathbf{b}
	=
	\frac{R_{0}^{3}}{M}\mathbf{B}
	=
	\frac{1}{R^{5}}
	\left(
	3XZ,3YZ,3Z^{2}-R^{2}
	\right),
	\label{eq:b_dimensionless}
\end{equation}
the Lorentz equation becomes
\begin{equation}
	\frac{d^{2}\mathbf{R}}{d\tau^{2}}
	=
	\frac{1}{\gamma}
	\frac{d\mathbf{R}}{d\tau}\times\mathbf{b}.
	\label{eq:cons_vector}
\end{equation}
Its Cartesian components are
\begin{align}
	\frac{d^{2}X}{d\tau^{2}}
	&=
	-\frac{1}{\gamma R^{3}}\frac{dY}{d\tau}
	+\frac{3Z}{\gamma R^{5}}
	\left(
	Z\frac{dY}{d\tau}
	-Y\frac{dZ}{d\tau}
	\right),
	\label{eq:cons_X}
	\\
	\frac{d^{2}Y}{d\tau^{2}}
	&=
	\frac{1}{\gamma R^{3}}\frac{dX}{d\tau}
	-\frac{3Z}{\gamma R^{5}}
	\left(
	Z\frac{dX}{d\tau}
	-X\frac{dZ}{d\tau}
	\right),
	\label{eq:cons_Y}
	\\
	\frac{d^{2}Z}{d\tau^{2}}
	&=
	\frac{3Z}{\gamma R^{5}}
	\left(
	Y\frac{dX}{d\tau}
	-X\frac{dY}{d\tau}
	\right).
	\label{eq:cons_Z}
\end{align}
These equations follow directly from the cross product in
Eq.~\eqref{eq:cons_vector}. They also show explicitly that initial data
satisfying $Z=0$ and $dZ/d\tau=0$ remain confined to the equatorial plane.

Writing
\begin{equation}
	\mathbf{V}
	=
	\frac{d\mathbf{R}}{d\tau},
	\qquad
	V^{2}=\mathbf{V}\cdot\mathbf{V},
\end{equation}
Eq.~\eqref{eq:cons_vector} gives
\begin{equation}
	\frac{dV^{2}}{d\tau}
	=
	2\mathbf{V}\cdot
	\frac{d\mathbf{V}}{d\tau}
	=
	0.
\end{equation}
Thus $V$ is constant, consistently with the absence of magnetic work, and
\begin{equation}
	\gamma
	=
	\frac{1}{\sqrt{1-\gamma_{0}V^{2}}},
	\qquad
	\gamma_{0}
	=
	\left(
	\frac{eM}{mc^{2}R_{0}^{2}}
	\right)^{2}.
	\label{eq:gamma_dimensionless}
\end{equation}
The physically relevant relativistic combination is $\gamma_{0}V^{2}=v^{2}/c^{2}$.
Consequently, the numerical value of the dimensionless speed $V$ alone
does not determine whether the motion is relativistic; the scaling
parameter $\gamma_{0}$ must also be specified.

For later use, the conserved dimensionless canonical angular momentum is
\begin{align}
	\mathcal{K}
	&\equiv
	\frac{cR_{0}}{eM}L_{z}
	\nonumber\\
	&=
	\gamma
	\left(
	X\frac{dY}{d\tau}
	-Y\frac{dX}{d\tau}
	\right)
	+
	\frac{X^{2}+Y^{2}}{R^{3}}.
	\label{eq:K_dimensionless}
\end{align}
The first term is the dimensionless mechanical angular momentum, whereas
the second is the magnetic contribution arising from the vector potential.
In the equatorial plane, $Z=0$, this reduces to
\begin{equation}
	\mathcal{K}
	=
	\gamma R^{2}\dot{\phi}
	+
	\frac{1}{R},
\end{equation}
where an overdot denotes differentiation with respect to $\tau$.

\subsection{Radiation reaction in the Landau--Lifshitz formulation}

Accelerated charges emit electromagnetic radiation, which carries energy and
momentum away from the particle. The associated recoil is described
classically by a radiation-reaction force. Even when this force is weak over
one orbital period, its cumulative effect may substantially modify the
long-term motion in a strongly magnetised and spatially inhomogeneous field.

The Lorentz--Abraham--Dirac equation is covariant, but its third-order
character permits runaway and pre-accelerating solutions. Landau and Lifshitz
instead apply reduction of order: the acceleration appearing in the
self-force is replaced by its Lorentz-force value, and terms are retained only
to first order in the radiation-reaction time
$\tau_{0}=2e^{2}/(3mc^{3})$, which for an electron is
$\tau_{0}\simeq6.26\times10^{-24}\,\mathrm{s}$. The resulting LL equation
is second order and applies when radiation reaction remains a perturbative
correction within the classical regime \cite{LL,DiPiazza:2011tq}.

For definiteness, let $x^{\mu}=(ct,\mathbf{x})$, let
$s=c\tau_{\rm p}$ denote proper length, define $u^{\mu}=dx^{\mu}/ds$, and
adopt the metric signature $(+,-,-,-)$, so that $u^{\mu}u_{\mu}=1$. In
Gaussian units, the LL equation can be written as
\begin{align}
	mc\,\frac{du^{\mu}}{ds}
	&=
	\frac{e}{c}F^{\mu\nu}u_{\nu}
	+
	\frac{2e^{3}}{3mc^{3}}
	\Biggl\{
	\bigl(\partial_{\alpha}F^{\mu\nu}\bigr)
	u^{\alpha}u_{\nu}
	\nonumber\\
	&\quad
	+
	\frac{e}{mc^{2}}
	F^{\mu\lambda}F_{\lambda\nu}u^{\nu}
	\nonumber\\
	&\quad
	+
	\frac{e}{mc^{2}}
	\bigl(F_{\nu\lambda}u^{\lambda}\bigr)
	\bigl(F^{\nu\rho}u_{\rho}\bigr)u^{\mu}
	\Biggr\}.
	\label{eq:LLcov}
\end{align}
The three terms inside braces contain, respectively, one derivative of the
external field and two contributions quadratic in the field tensor. The
first is sensitive to spatial inhomogeneity, whereas the latter two govern
the dissipative energy balance.

For the static magnetic field considered here, with vanishing electric
field, the spatial LL force added to the right-hand side of
Eq.~\eqref{eq:Lorentz} is
\begin{align}
	\mathbf{F}_{\rm LL}
	&=
	\frac{2e^{3}}{3mc^{4}}\,
	\gamma\,
	\mathbf{v}\times
	\left[
	(\mathbf{v}\!\cdot\!\boldsymbol{\nabla})\mathbf{B}
	\right]
	\nonumber\\
	&\quad
	+
	\frac{2e^{4}}{3m^{2}c^{5}}\,
	\mathbf{B}\times
	(\mathbf{B}\times\mathbf{v})
	\nonumber\\
	&\quad
	-
	\frac{2e^{4}}{3m^{2}c^{7}}\,
	\gamma^{2}\mathbf{v}
	\left[
	B^{2}v^{2}
	-
	(\mathbf{v}\!\cdot\!\mathbf{B})^{2}
	\right].
	\label{eq:LL3D}
\end{align}
The first line is the field-gradient contribution. It is not the total time
derivative of $\gamma(\mathbf{v}\times\mathbf{B})$. Denoting it by
$\mathbf{F}_{\rm grad}$, one has
$\mathbf{v}\cdot\mathbf{F}_{\rm grad}=0$; it therefore performs no work,
although it may deflect the momentum and modify the local orbit. The
quadratic-field terms provide the radiative damping.

Taking the scalar product of Eq.~\eqref{eq:LL3D} with $\mathbf{v}$ gives
\begin{align}
	\frac{d}{dt}(\gamma mc^{2})
	&=
	-
	\frac{2e^{4}}{3m^{2}c^{5}}\,
	\gamma^{2}
	\left[
	B^{2}v^{2}
	-
	(\mathbf{v}\!\cdot\!\mathbf{B})^{2}
	\right]
	\leq 0.
	\label{eq:LL_energy_loss}
\end{align}
The quantity in square brackets is $B^{2}v_{\perp}^{2}$, where
$v_{\perp}$ is the velocity perpendicular to the magnetic field. Thus the
energy loss is controlled by the component of the motion accelerated by the
Lorentz force. The field-gradient term does not enter this instantaneous
energy balance, but it cannot generally be neglected in the momentum
dynamics.

For motion confined to the equatorial plane of the dipole,
$\mathbf{v}\cdot\mathbf{B}=0$. If the field-gradient term is omitted, the
two quadratic-field contributions combine, using
$1+\gamma^{2}v^{2}/c^{2}=\gamma^{2}$, to give
\begin{equation}
	\mathbf{F}_{\rm rad}^{\rm (eff)}
	=
	-
	\frac{2e^{4}}{3m^{2}c^{5}}\,
	\gamma^{2}B^{2}\mathbf{v}.
	\label{eq:FradEffective}
\end{equation}
For an exactly circular equatorial orbit,
$(\mathbf{v}\cdot\boldsymbol{\nabla})\mathbf{B}=0$, and
Eq.~\eqref{eq:FradEffective} is therefore the complete LL correction. For a
noncircular equatorial orbit, the field-gradient term need not vanish, even
though it still performs no work.

In the remainder of this work, Eq.~\eqref{eq:FradEffective} and its
three-dimensional quadratic-field generalisation are used as a reduced
drag-only model. This approximation reproduces the exact LL energy-loss law
in a static magnetic field while omitting directional effects associated
with field inhomogeneity. Its range of validity near the circular equatorial
branch will be assessed separately.

\subsection{Dimensionless equations of motion with self-force}

We now express the LL dynamics in the dimensionless variables introduced
above. For the dipole field in Eq.~\eqref{eq:b_dimensionless},
\begin{equation}
	b^{2}
	=
	\frac{R^{2}+3Z^{2}}{R^{8}}.
	\label{eq:b_squared}
\end{equation}
Thus the field strength scales as $R^{-6}$, with an additional angular
dependence away from the equatorial plane. At $Z=0$, one has
$\mathbf{b}=(0,0,-R^{-3})$ and $b^{2}=R^{-6}$.

Radiation reaction introduces the dimensionless parameter
\begin{equation}
	\eta
	=
	\frac{2e^{3}M}{3m^{2}c^{4}R_{0}^{3}}
	=
	\frac{\tau_{0}}{t_{0}},
	\label{eq:eta}
\end{equation}
where $t_{0}=mcR_{0}^{3}/(eM)$ is the characteristic St\"{o}rmer time and
$eM>0$ is assumed. Hence $\eta$ measures the separation between the
radiation-reaction and orbital timescales. For an electron with
$M=10^{25}\,{\rm G\,cm^{3}}$ and $R_{0}=10^{10}\,{\rm cm}$, one obtains
$\eta\simeq1.10\times10^{-21}$ using $|e|$. Although this value is very
small, the local damping is enhanced in the inner dipole region by the rapid
growth of $b^{2}$.

Since $\gamma$ is no longer constant, it is convenient to introduce the
dimensionless relativistic momentum
\begin{equation}
	\mathbf{p}
	=
	\gamma\mathbf{V},
	\qquad
	\gamma
	=
	\sqrt{1+\gamma_{0}p^{2}},
	\label{eq:p_gamma_relation}
\end{equation}
where $\mathbf{V}=d\mathbf{R}/d\tau$ and $\gamma_{0}$ is defined in
Eq.~\eqref{eq:gamma_dimensionless}. The complete dimensionless LL system is
\begin{equation}
	\begin{aligned}
		\frac{d\mathbf{R}}{d\tau}
		&=
		\mathbf{V}
		=
		\frac{\mathbf{p}}{\gamma},
		\\[0.5ex]
		\frac{d\mathbf{p}}{d\tau}
		&=
		\mathbf{V}\times\mathbf{b}
		+
		\mathbf{A}_{\rm rad}^{\rm(full)}.
	\end{aligned}
	\label{eq:mom_eq}
\end{equation}
The complete dimensionless radiation-reaction term is
\begin{align}
	\mathbf{A}_{\rm rad}^{\rm(full)}
	&=
	\eta\gamma\,
	\mathbf{V}\times
	\left[
	(\mathbf{V}\!\cdot\!\boldsymbol{\nabla}_{\!R})
	\mathbf{b}
	\right]
	\nonumber\\
	&\quad
	+
	\eta\left[
	(\mathbf{V}\!\cdot\!\mathbf{b})\mathbf{b}
	-b^{2}\mathbf{V}
	\right]
	\nonumber\\
	&\quad
	-
	\eta\gamma_{0}\gamma^{2}\mathbf{V}
	\left[
	b^{2}V^{2}
	-(\mathbf{V}\!\cdot\!\mathbf{b})^{2}
	\right].
	\label{eq:Arad_full}
\end{align}
Here $\boldsymbol{\nabla}_{\!R}$ denotes differentiation with respect to the
dimensionless coordinates. The first term is the field-gradient contribution
and is the direct dimensionless counterpart of the first line of
Eq.~\eqref{eq:LL3D}; it is not a total derivative of
$\mathbf{p}\times\mathbf{b}$. The first quadratic-field term is
$-\eta b^{2}\mathbf{V}_{\perp}$, where $\mathbf{V}_{\perp}$ is the velocity
perpendicular to $\mathbf{b}$, while the last term supplies the additional
relativistic damping along $\mathbf{V}$.

The field-gradient term is orthogonal to $\mathbf{V}$ and therefore performs
no work. The complete dimensionless LL equation consequently gives
\begin{align}
	\frac{d\gamma}{d\tau}
	&=
	-\eta\gamma_{0}\gamma^{2}
	\left[
	b^{2}V^{2}
	-(\mathbf{V}\!\cdot\!\mathbf{b})^{2}
	\right]
	\leq 0.
	\label{eq:dimensionless_energy_loss}
\end{align}
The quantity in square brackets is $b^{2}V_{\perp}^{2}$. Thus the particle
energy is nonincreasing, but this fact alone does not determine whether a
general orbit moves inward or outward.

For the planar analytical and numerical treatment below, we use the reduced
model obtained by omitting the field-gradient term while retaining both
quadratic-field contributions. Its radiation term may be written in the two
equivalent forms
\begin{eqnarray}
	\mathbf{A}_{\rm rad}^{\rm(eff)}
	&=&
	\eta\left[
	(\mathbf{V}\!\cdot\!\mathbf{b})\mathbf{b}
	-b^{2}\mathbf{V}
	\right]
	\nonumber\\
	&&\quad
	-
	\eta\gamma_{0}\gamma^{2}\mathbf{V}
	\left[
	b^{2}V^{2}
	-(\mathbf{V}\!\cdot\!\mathbf{b})^{2}
	\right],
	\label{eq:Arad_eff}
		\nonumber \\
	&=&
	\eta\Bigl[
	\mathbf{b}\times
	(\mathbf{b}\times\mathbf{V})
	\nonumber\\
	&&\quad
	-
	\gamma_{0}\gamma^{2}\mathbf{V}
	\left[
	b^{2}V^{2}
	-(\mathbf{V}\!\cdot\!\mathbf{b})^{2}
	\right]
	\Bigr].
	\label{eq:Arad_eff2}
\end{eqnarray}
This is a local drag-only approximation, not an orbital average. Because the
omitted term is orthogonal to $\mathbf{V}$, the reduced model preserves the
exact energy-loss law in Eq.~\eqref{eq:dimensionless_energy_loss}, while
omitting directional effects caused by field inhomogeneity.

The reduced system is therefore
\begin{equation}
	\begin{aligned}
		\frac{d\mathbf{R}}{d\tau}
		&=
		\frac{\mathbf{p}}{\gamma},
		\\[0.5ex]
		\frac{d\mathbf{p}}{d\tau}
		&=
		\frac{\mathbf{p}}{\gamma}\times\mathbf{b}
		+
		\mathbf{A}_{\rm rad}^{\rm(eff)},
		\\[0.5ex]
		\gamma
		&=
		\sqrt{1+\gamma_{0}p^{2}}.
	\end{aligned}
	\label{eq:sys_final}
\end{equation}
Unlike the conservative CRSP, this system does not conserve the particle
energy or, in general, the canonical angular momentum.

For strictly equatorial motion,
$\mathbf{V}\cdot\mathbf{b}=0$ and $b^{2}=R^{-6}$. Using
$1+\gamma_{0}\gamma^{2}V^{2}=\gamma^{2}$, the reduced radiation term becomes
\begin{equation}
	\left.
	\mathbf{A}_{\rm rad}^{\rm(eff)}
	\right|_{\rm eq}
	=
	-\eta\gamma^{2}b^{2}\mathbf{V}
	=
	-\eta\gamma b^{2}\mathbf{p}.
	\label{eq:Arad_eq}
\end{equation}
The equatorial momentum equation is consequently
\begin{equation}
	\frac{d\mathbf{p}}{d\tau}
	=
	\frac{\mathbf{p}}{\gamma}\times\mathbf{b}
	-
	\eta\gamma b^{2}\mathbf{p}.
	\label{eq:sys_eq}
\end{equation}
For an exactly circular equatorial orbit, the omitted field-gradient term
vanishes, so the reduced and complete LL corrections coincide. For
noncircular equatorial or genuinely three-dimensional motion, the reduced
system remains an explicit modelling approximation whose validity must be
assessed separately.

\section{Planar motion with radiation reaction}
\label{sec:planar}

A particular example of motion in a dipole magnetic field, with important astrophysical applications,  is the evolution of a charged particle in the equatorial plane of a massive object. In the following We investigate in detail  the main physical properties of the planar St{\o}rmer problem in the presence of radiation reaction. In particular, the exact solution of the motion along a circular branch is also presented.    

\subsection{Adiabatic evolution along the circular branch}
\label{sec:planar_decay}

In the conservative CRSP, a moving circular orbit in the equatorial plane,
$Z=0$, is an exact solution. For the orientation fixed by $eM>0$, its
positive angular frequency and speed satisfy
\begin{equation}
	\omega
	=
	\frac{1}{\gamma R^{3}},
	\qquad
	V
	=
	\omega R
	=
	\frac{1}{\gamma R^{2}}.
	\label{eq:circ_prop}
\end{equation}
Since $\mathbf{p}=\gamma\mathbf{V}$, the dimensionless mechanical angular
momentum is
\begin{equation}
	\ell_{\rm mech}
	=
	Xp_{Y}-Yp_{X}
	=
	\gamma R^{2}\dot{\phi}
	=
	\frac{1}{R}.
	\label{eq:lmech_circular}
\end{equation}
The canonical angular momentum is therefore $\mathcal{K}=2/R$. Combining
Eq.~\eqref{eq:circ_prop} with the definition of the Lorentz factor gives
\begin{equation}
	\gamma^{2}
	=
	1+\frac{\gamma_{0}}{R^{4}}.
	\label{eq:gamma_circular}
\end{equation}
Thus smaller circular orbits are more relativistic, whereas the branch
approaches the nonrelativistic regime as $R$ increases.

The moving circular branch is radially unstable. At fixed $\gamma$ and
$\mathcal{K}$, the planar radial first integral may be written as
$\dot R^{2}+U(R)=V^{2}$, with
$U(R)=(\mathcal{K}-R^{-1})^{2}/(\gamma^{2}R^{2})$. The moving circular
solution corresponds to $\mathcal{K}=2/R_{c}$ and satisfies
\begin{equation}
	U''(R_{c})
	=
	-\frac{4}{\gamma^{2}R_{c}^{6}}
	<0.
	\label{eq:circular_radial_instability}
\end{equation}
The alternative stationary condition $\mathcal{K}=1/R_{c}$ has vanishing
mechanical angular momentum and does not represent the moving circular
branch. A radial perturbation of the latter grows at the rate
$\sqrt{2}/(\gamma R_{c}^{3})=\sqrt{2}\,\omega$, namely, on the orbital
rather than the dissipative timescale. The evolution derived below should
therefore be interpreted as a conditional secular solution constrained to
follow the instantaneous circular branch, and as an analytical benchmark
rather than the generic fate of a nearby orbit.

For the reduced equatorial system in Eq.~\eqref{eq:sys_eq}, the Lorentz term
rotates the momentum without changing its magnitude. Taking the scalar
product of the momentum equation with $\mathbf{p}$ gives
\begin{equation}
	\frac{1}{2}\frac{dp^{2}}{d\tau}
	=
	-\eta\,\frac{\gamma}{R^{6}}\,p^{2}.
	\label{eq:p2loss}
\end{equation}
Using $p^{2}=(\gamma^{2}-1)/\gamma_{0}$, one obtains
\begin{equation}
	\frac{d\gamma}{d\tau}
	=
	-\eta\,\frac{\gamma^{2}-1}{R^{6}}.
	\label{eq:dg}
\end{equation}
The factor $R^{-6}$ reflects the strong localisation of radiative losses in
the inner dipole region. On the circular branch,
Eq.~\eqref{eq:gamma_circular} reduces this law to
\begin{equation}
	\frac{d\gamma}{d\tau}
	=
	-\eta\,\frac{\gamma_{0}}{R^{10}}.
	\label{eq:dg_circ}
\end{equation}

Differentiating Eq.~\eqref{eq:gamma_circular} along the slowly evolving
branch yields
\begin{equation}
	\frac{dR}{d\tau}
	=
	-\frac{\gamma R^{5}}{2\gamma_{0}}
	\frac{d\gamma}{d\tau}.
	\label{eq:dR_dgamma}
\end{equation}
Substitution of Eq.~\eqref{eq:dg_circ} then gives
\begin{equation}
	\frac{dR}{d\tau}
	=
	\frac{\eta\gamma}{2R^{5}}.
	\label{eq:dRdt}
\end{equation}
Hence energy loss is accompanied by outward motion along this particular
branch. This behaviour is physically possible because the circular speed
required by magnetic force balance decreases with radius: as radiation
reduces the momentum, the corresponding instantaneous circular state lies
farther from the dipole. It does not follow that arbitrary planar
trajectories migrate outward.

In the nonrelativistic regime, $R\gg\gamma_{0}^{1/4}$ and $\gamma\simeq1$.
Equation~\eqref{eq:dRdt} then gives
\begin{equation}
	R^{6}(\tau)
	\simeq
	R_{i}^{6}+3\eta(\tau-\tau_{i}),
	\label{eq:R_late_time}
\end{equation}
where $R_{i}=R(\tau_{i})$. Thus $R\propto\tau^{1/6}$ at late times. The
drift becomes progressively slower because both the dipole field and the
radiative loss decrease rapidly with radius.

Along the same branch,
\begin{equation}
	\begin{aligned}
		\omega
		&=
		\frac{1}{R\sqrt{R^{4}+\gamma_{0}}},
		\\[0.5ex]
		\gamma^{2}b^{2}V^{2}
		&=
		\frac{1}{R^{10}}.
	\end{aligned}
	\label{eq:frequency_power_scaling}
\end{equation}
The orbital frequency and the kinematic-field factor entering the LL
energy-loss law both decrease monotonically with $R$. The branch therefore
has no finite terminal radius: formally, $R\to\infty$,
$dR/d\tau\to0$, and $\gamma\to1$ only as $\tau\to\infty$.

\subsection{Validity of the drag-dominated approximation}
\label{sec:gradient_terms}

The reduced model used in the preceding subsection omits the field-gradient
term of the complete Landau--Lifshitz force. This is an approximation to the
local momentum dynamics, not the removal of a total derivative. From
Eq.~\eqref{eq:Arad_full}, the neglected contribution is
\begin{equation}
	\mathbf{A}_{\rm grad}
	=
	\eta\gamma
	\mathbf{V}\times
	\left[
	(\mathbf{V}\cdot\boldsymbol{\nabla}_{R})
	\mathbf{b}
	\right].
	\label{eq:Arad_grad}
\end{equation}
Since $\mathbf{V}\cdot\mathbf{A}_{\rm grad}=0$, this term performs no work
and does not affect the instantaneous LL energy-loss law. It may nevertheless
deflect the momentum and thereby alter the local trajectory. Consequently,
the reduced model reproduces the exact evolution of the particle energy in a
static magnetic field, but not necessarily its complete spatial motion.

In the equatorial plane,
$\mathbf{b}=-R^{-3}\mathbf{e}_{z}$ and $b^{2}=R^{-6}$. Writing
$\mathbf{V}=V_{R}\mathbf{e}_{R}+V_{\phi}\mathbf{e}_{\phi}$ gives
\begin{equation}
	(\mathbf{V}\cdot\boldsymbol{\nabla}_{R})
	\mathbf{b}
	=
	\frac{3V_{R}}{R^{4}}
	\mathbf{e}_{z}.
\end{equation}
The field-gradient term therefore vanishes identically on an exactly
circular equatorial orbit, for which $V_{R}=0$. For nearly circular motion,
its magnitude is
\begin{equation}
	\left|\mathbf{A}_{\rm grad}\right|
	=
	\frac{3\eta\gamma}{R^{4}}
	\left|V_{R}\right|V.
	\label{eq:grad_magnitude}
\end{equation}

The radiation term retained in the reduced equatorial model is
\begin{equation}
	\mathbf{A}_{\rm drag}
	=
	-\eta\gamma^{2}
	\frac{\mathbf{V}}{R^{6}}
	=
	-\eta
	\frac{\gamma}{R^{6}}
	\mathbf{p}.
	\label{eq:Arad_eff_planar}
\end{equation}
The exact ratio of the two magnitudes is therefore
\begin{equation}
	\frac{
		\left|\mathbf{A}_{\rm grad}\right|
	}{
		\left|\mathbf{A}_{\rm drag}\right|
	}
	=
	\frac{3R^{2}\left|V_{R}\right|}{\gamma}.
	\label{eq:ratio_grad_drag}
\end{equation}
Both terms are proportional to $\eta$, so the relative accuracy of the
drag-only approximation is independent of the absolute strength of radiation
reaction.

Near the circular branch,
$V\simeq\left|V_{\phi}\right|\simeq1/(\gamma R^{2})$, and the ratio becomes
\begin{equation}
	\frac{
		\left|\mathbf{A}_{\rm grad}\right|
	}{
		\left|\mathbf{A}_{\rm drag}\right|
	}
	\simeq
	\frac{3}{\gamma^{2}}
	\frac{
		\left|V_{R}\right|
	}{
		\left|V_{\phi}\right|
	}.
	\label{eq:ratio_grad_drag_circular}
\end{equation}
The reduced model is therefore locally controlled when
\begin{equation}
	\frac{
		\left|V_{R}\right|
	}{
		\left|V_{\phi}\right|
	}
	\ll
	\frac{\gamma^{2}}{3}.
	\label{eq:drag_condition}
\end{equation}
The relevant small parameter is the radial fraction of the velocity. For a
fixed value of $\left|V_{R}\right|/\left|V_{\phi}\right|$, the approximation
also improves as $\gamma$ increases.

On the circular branch itself, the reduced and complete LL corrections
coincide exactly. A nearby orbit, however, need not remain in this regime.
The radial instability discussed in Sec.~\ref{sec:planar_decay} can amplify
$V_{R}$ on the orbital timescale and eventually invalidate
Eq.~\eqref{eq:drag_condition}. The approximation should therefore be viewed
as a local description of nearly circular equatorial motion rather than as a
uniform approximation throughout the planar phase space.

For eccentric equatorial or genuinely three-dimensional trajectories, the
field-gradient term may become comparable to the quadratic-field terms even
though it continues to perform no work. Such trajectories require the
complete LL momentum equation whenever directional accuracy is important.
Within the restricted circular-branch analysis, however, the evolution in
Eq.~\eqref{eq:dRdt} is unaffected by the omitted term because
$\mathbf{A}_{\rm grad}=0$ on the branch itself.

\subsection{Exact analytical solution along the circular branch}
\label{sec:exact_circ}

The conditional circular-branch evolution in Eq.~\eqref{eq:dRdt} can be
integrated exactly. Using
$\gamma=\sqrt{R^{4}+\gamma_{0}}/R^{2}$, one obtains
\begin{equation}
	\frac{d\tau}{dR}=
	\frac{2R^{5}}{\eta\gamma}
	=
	\frac{2R^{7}}
	{\eta\sqrt{R^{4}+\gamma_{0}}}.
	\label{eq:dtau_dR}
\end{equation}
For an initial condition $R(0)=R_{i}$, separation of variables gives
\begin{equation}
	\tau
	=
	\frac{2}{\eta}
	\int_{R_{i}}^{R}
	\frac{\rho^{7}}
	{\sqrt{\rho^{4}+\gamma_{0}}}
	\mathrm{d}\rho.
\end{equation}
Introducing $u=\rho^{2}$ reduces the antiderivative to
\begin{align}
	\int
	\frac{u^{3}}
	{\sqrt{u^{2}+\gamma_{0}}}
	\mathrm{d}u
	&=
	\frac{1}{3}
	\left(u^{2}+\gamma_{0}\right)^{3/2}
	\nonumber\\
	&\quad
	-
	\gamma_{0}
	\left(u^{2}+\gamma_{0}\right)^{1/2}.
\end{align}
It is therefore convenient to define
\begin{equation}
	\mathcal{F}(R)
	\equiv
	\frac{1}{3}
	\left(R^{4}+\gamma_{0}\right)^{3/2}
	-
	\gamma_{0}
	\left(R^{4}+\gamma_{0}\right)^{1/2}.
	\label{eq:F_exact}
\end{equation}
The exact implicit solution is then
\begin{equation}
	\tau(R)
	=
	\frac{1}{\eta}
	\left[
	\mathcal{F}(R)-\mathcal{F}(R_{i})
	\right].
	\label{eq:tau_exact}
\end{equation}
Since
$\mathcal{F}'(R)=2R^{7}/\sqrt{R^{4}+\gamma_{0}}>0$ for $R>0$,
this relation defines a unique radius that increases monotonically with
$\tau$ along the circular branch. In practice, Eq.~\eqref{eq:tau_exact}
may be inverted numerically without integrating the equations of motion.

In the nonrelativistic regime, $R^{4}\gg\gamma_{0}$, the function
$\mathcal{F}$ has the expansion
\begin{equation}
	\mathcal{F}(R)
	=
	\frac{R^{6}}{3}
	-
	\frac{\gamma_{0}R^{2}}{2}
	-
	\frac{3\gamma_{0}^{2}}{8R^{2}}
	+
	\mathcal{O}\left(
	\frac{\gamma_{0}^{3}}{R^{6}}
	\right).
\end{equation}
When both $R$ and $R_{i}$ lie in this regime, the leading evolution is
\begin{equation}
	\tau
	\simeq
	\frac{R^{6}-R_{i}^{6}}{3\eta},
\end{equation}
which recovers the late-time scaling $R\propto\tau^{1/6}$.

In the ultrarelativistic regime, $R^{4}\ll\gamma_{0}$,
\begin{equation}
	\mathcal{F}(R)
	=
	-\frac{2}{3}\gamma_{0}^{3/2}
	+
	\frac{R^{8}}{4\sqrt{\gamma_{0}}}
	-
	\frac{R^{12}}{12\gamma_{0}^{3/2}}
	+
	\mathcal{O}\left(
	\frac{R^{16}}{\gamma_{0}^{5/2}}
	\right).
\end{equation}
If both radii remain within this regime, the leading evolution is
\begin{equation}
	\tau
	\simeq
	\frac{R^{8}-R_{i}^{8}}
	{4\eta\sqrt{\gamma_{0}}},
\end{equation}
corresponding to $R\propto\tau^{1/8}$. The crossover between the two
asymptotic behaviours occurs at radii of order $\gamma_{0}^{1/4}$, where the
motion changes from relativistic to nonrelativistic along the branch.

The orbital frequency is
$\omega(R)=1/[R\sqrt{R^{4}+\gamma_{0}}]$, while the kinematic-field factor
entering the LL energy-loss law is
$\gamma^{2}b^{2}V^{2}=R^{-10}$. Both decrease monotonically as the radius
grows. The exact solution therefore provides a useful benchmark for numerical
integrations and clearly displays the progressive weakening of the radiative
evolution. These orbital scalings do not, by themselves, determine the
emitted spectrum. Moreover, Eqs.~\eqref{eq:F_exact} and
\eqref{eq:tau_exact} apply only to motion constrained to the instantaneous
circular branch and do not imply that a generic nearby orbit remains on that
radially unstable branch.

\subsection{Averaged phase-space transport and radial action}
\label{sec:adiab_inv}

For conservative equatorial motion, the Lorentz factor $\gamma$ and the
canonical angular momentum
\begin{equation}
	\mathcal{K}
	=
	\gamma R^{2}\dot{\phi}
	+
	\frac{1}{R}
	\label{eq:K_def}
\end{equation}
are constant. The radial dynamics is governed by
\begin{equation}
	\dot{R}^{2}
	+
	\frac{\left(\mathcal{K}-R^{-1}\right)^{2}}
	{\gamma^{2}R^{2}}
	=
	V^{2},
	\label{eq:radial_int}
\end{equation}
where
$V^{2}=(\gamma^{2}-1)/(\gamma_{0}\gamma^{2})$.
For suitable values of $(\gamma,\mathcal{K})$, this equation admits regular
bound librations between the turning points $R_{\min}$ and $R_{\max}$.

For the reduced equatorial model in Eq.~\eqref{eq:sys_eq}, the exact
instantaneous drift equations are
\begin{align}
	\frac{\mathrm{d}\gamma}{\mathrm{d}\tau}
	&=
	-\eta
	\frac{\gamma^{2}-1}{R^{6}},
	\label{eq:dg_exact}
	\\
	\frac{\mathrm{d}\mathcal{K}}{\mathrm{d}\tau}
	&=
	-\eta
	\frac{\gamma}{R^{6}}
	\left(
	\mathcal{K}-\frac{1}{R}
	\right).
	\label{eq:dK_exact}
\end{align}
The factor $R^{-6}$ makes the dissipative evolution particularly sensitive
to passages through the inner, high-field part of the orbit. Moreover,
$\mathcal{K}-1/R$ is the mechanical angular momentum. Radiation reaction
therefore decreases $\mathcal{K}$ for positive mechanical angular momentum
and increases it when the mechanical angular momentum is negative.

When the fractional dissipative change during one conservative period is
small, one may introduce the slow time $T=\eta\tau$ and average over the
corresponding conservative libration. The orbit must also remain sufficiently
far from a separatrix, where the conservative period becomes large and the
separation of timescales may fail. The averaged equations are
\begin{equation}
	\begin{aligned}
		\frac{\mathrm{d}\gamma}{\mathrm{d}T}
		&=
		-\left\langle
		\frac{\gamma^{2}-1}{R^{6}}
		\right\rangle,
		\\[0.5ex]
		\frac{\mathrm{d}\mathcal{K}}{\mathrm{d}T}
		&=
		-\left\langle
		\frac{\gamma}{R^{6}}
		\left(
		\mathcal{K}-\frac{1}{R}
		\right)
		\right\rangle.
	\end{aligned}
	\label{eq:slow_gammaK}
\end{equation}
For any function $f(R)$, the conservative orbital average is
\begin{equation}
	\begin{aligned}
		\langle f\rangle
		&=
		\frac{2}{T_{\rm orb}}
		\int_{R_{\min}}^{R_{\max}}
		\frac{f(R)}
		{\left|\dot{R}\right|}
		\,\mathrm{d}R,
		\\[0.5ex]
		T_{\rm orb}
		&=
		2
		\int_{R_{\min}}^{R_{\max}}
		\frac{\mathrm{d}R}
		{\left|\dot{R}\right|}.
	\end{aligned}
	\label{eq:orbital_average}
\end{equation}
where $\left|\dot R\right|$ follows from
Eq.~\eqref{eq:radial_int}. During each average, $\gamma$ and
$\mathcal{K}$ are treated as fixed labels of the underlying conservative
orbit.

The moving circular branch discussed in Sec.~\ref{sec:planar_decay} is not
the regular small-amplitude limit of these librations because it corresponds
to a maximum of the radial effective potential. Its conditional evolution
must therefore be treated separately.

The averages in Eq.~\eqref{eq:slow_gammaK} are functions of the instantaneous
conservative orbit and hence of $(\gamma,\mathcal{K})$. They do not,
however, reduce to a universal algebraic relation between the two drift
rates. No general conserved combination of $\gamma$ and $\mathcal{K}$ follows
from the averaged system, and their secular evolution must be determined
jointly.

A useful measure of the radial libration is the canonical radial action. The
radial canonical momentum is $p_{R}=\gamma\dot{R}$ and satisfies
\begin{equation}
	p_{R}^{2}
	=
	\frac{\gamma^{2}-1}{\gamma_{0}}
	-
	\frac{\left(\mathcal{K}-R^{-1}\right)^{2}}
	{R^{2}}.
	\label{eq:pR_radial}
\end{equation}
The corresponding action is
\begin{align}
	J_{R}
	&\equiv
	\frac{1}{2\pi}
	\oint p_{R}\mathrm{d}R
	\nonumber\\
	&=
	\frac{1}{\pi}
	\int_{R_{\min}}^{R_{\max}}
	\left[
	\frac{\gamma^{2}-1}{\gamma_{0}}
	-
	\frac{\left(\mathcal{K}-R^{-1}\right)^{2}}
	{R^{2}}
	\right]^{1/2}
	\mathrm{d}R.
	\label{eq:J_R}
\end{align}
The expression
$\oint\dot{R}\mathrm{d}R/(2\pi)$ is not the canonical action because it
omits the relativistic factor $\gamma$.

Along the averaged dissipative evolution, the action changes according to
\begin{equation}
	\frac{\mathrm{d}J_{R}}{\mathrm{d}T} =
	\frac{\partial J_{R}}{\partial\gamma}
	\frac{\mathrm{d}\gamma}{\mathrm{d}T}
	+
	\frac{\partial J_{R}}{\partial\mathcal{K}}
	\frac{\mathrm{d}\mathcal{K}}{\mathrm{d}T}.
	\label{eq:J_invariant}
\end{equation}
The endpoint contributions generated when differentiating the action vanish
because $p_{R}=0$ at the turning points. There is no general cancellation
between the two terms in Eq.~\eqref{eq:J_invariant}. Thus $J_{R}$ is a useful
phase-space diagnostic, but its conservation is not implied by the adiabatic
approximation. Whether the radial libration grows or decays must be determined
from the averaged drift equations or from direct integration of the reduced
dynamical system.

A finite terminal state cannot be obtained by assuming conservation of
$J_{R}$. Indeed,
\begin{equation}
	0
	\leq
	p_{R}^{2}
	\leq
	\frac{\gamma^{2}-1}{\gamma_{0}}
	\longrightarrow
	0
	\qquad
	\text{as}
	\qquad
	\gamma\longrightarrow1.
	\label{eq:J_freeze}
\end{equation}
Therefore, for any family of bound librations whose turning points remain
finite as $\gamma\to1$, the radial action also approaches zero. A nonzero
initial action cannot remain constant in such a limit. More generally, the
present analysis predicts neither a universal terminal radius nor generic
circularisation: planar phase-space transport is controlled by the coupled
drift of $\gamma$ and $\mathcal{K}$.

\subsection{Numerical evolution in the invariant equatorial plane}
\label{sec:rr_numerical}

We now illustrate the planar dynamics of the reduced system introduced in
Eq.~\eqref{eq:sys_final}. The field-gradient contribution of the complete
Landau--Lifshitz force is omitted, whereas both terms quadratic in the
magnetic field are retained. The trajectories presented below are therefore
solutions of the reduced drag-only model and not of the complete local LL
equation. We take $\eta>0$, consistently with the convention $eM>0$.

Initial data satisfying $Z=p_{Z}=0$ remain in the equatorial plane. This
invariance follows from reflection symmetry about $Z=0$: neither the
equatorial Lorentz force nor the reduced radiation term generates a vertical
momentum component. With
$\mathbf{b}=(0,0,-R^{-3})$ and
$R=(X^{2}+Y^{2})^{1/2}$, Eq.~\eqref{eq:sys_eq} becomes
\begin{align}
	\dot{X}
	&=
	\frac{p_{X}}{\gamma},
	\qquad \qquad
	\dot{Y}
	=
	\frac{p_{Y}}{\gamma},
	\label{eq:rr_planar_positions}
	\\
	\dot{p}_{X}
	&=
	-\frac{p_{Y}}{\gamma R^{3}}
	-
	\eta\frac{\gamma}{R^{6}}p_{X},
	\nonumber\\
	\dot{p}_{Y}
	&=
	\frac{p_{X}}{\gamma R^{3}}
	-
	\eta\frac{\gamma}{R^{6}}p_{Y}.
	\label{eq:rr_planar_momenta}
\end{align}
Here an overdot denotes differentiation with respect to $\tau$, and
\begin{equation}
	\gamma
	=
	\sqrt{
		1+\gamma_{0}
		\left(
		p_{X}^{2}+p_{Y}^{2}
		\right)
	}.
	\label{eq:rr_planar_gamma}
\end{equation}
The first terms in Eq.~\eqref{eq:rr_planar_momenta} rotate the momentum
through the Lorentz force, whereas the terms proportional to $\eta$ reduce
its magnitude. Because their coefficient varies as $R^{-6}$, the damping is
strongly concentrated near the dipole.

Two exact relations provide useful checks on the numerical integration.
First,
\begin{equation}
	\frac{1}{2}
	\frac{\mathrm{d}p^{2}}{\mathrm{d}\tau}
	=
	-\eta
	\frac{\gamma}{R^{6}}
	p^{2}.
	\label{eq:rr_p2_decay}
\end{equation}
Integration gives
\begin{align}
	\ln
	\left[
	\frac{p^{2}(\tau)}
	{p^{2}(0)}
	\right]
	&=
	-2\eta
	\int_{0}^{\tau}
	\frac{\gamma(\tau')}
	{R^{6}(\tau')}
	\mathrm{d}\tau'.
	\label{eq:rr_p2_integral}
\end{align}
Thus the momentum does not decay exponentially with a constant coefficient.
The instantaneous logarithmic damping rate is
$2\eta\gamma/R^{6}$ and changes continuously as the particle moves through
regions of different field strength. Repeated passages through small radii
can therefore dominate the accumulated energy loss even when the particle
spends most of its time farther from the dipole.

Equivalently, the Lorentz factor satisfies
\begin{equation}
	\frac{\mathrm{d}\gamma}{\mathrm{d}\tau}
	=
	-\eta
	\frac{\gamma^{2}-1}{R^{6}}
	\leq 0.
	\label{eq:rr_num_energy}
\end{equation}
The particle energy is consequently monotonic, but the radius need not be.
Energy loss constrains the motion in momentum space; it does not by itself
select an inward or outward direction in configuration space.

The second diagnostic concerns the canonical angular momentum,
\begin{equation}
	\mathcal{K}
	=
	Xp_{Y}
	-
	Yp_{X}
	+
	\frac{1}{R}.
	\label{eq:rr_num_K}
\end{equation}
Its evolution is
\begin{align}
	\frac{\mathrm{d}\mathcal{K}}{\mathrm{d}\tau}
	&=
	-\eta
	\frac{\gamma}{R^{6}}
	\left(
	\mathcal{K}
	-
	\frac{1}{R}
	\right).
	\label{eq:rr_num_K_decay}
\end{align}
The quantity in parentheses is the mechanical angular momentum
$Xp_{Y}-Yp_{X}$. Hence $\mathcal{K}$ decreases for counterclockwise motion
with positive mechanical angular momentum and increases for clockwise motion
with negative mechanical angular momentum. Radiation reaction therefore does
not impose a universal sign on the canonical-angular-momentum drift.

\begin{figure*}[t]
	\centering
	\includegraphics[width=0.48\textwidth]
	{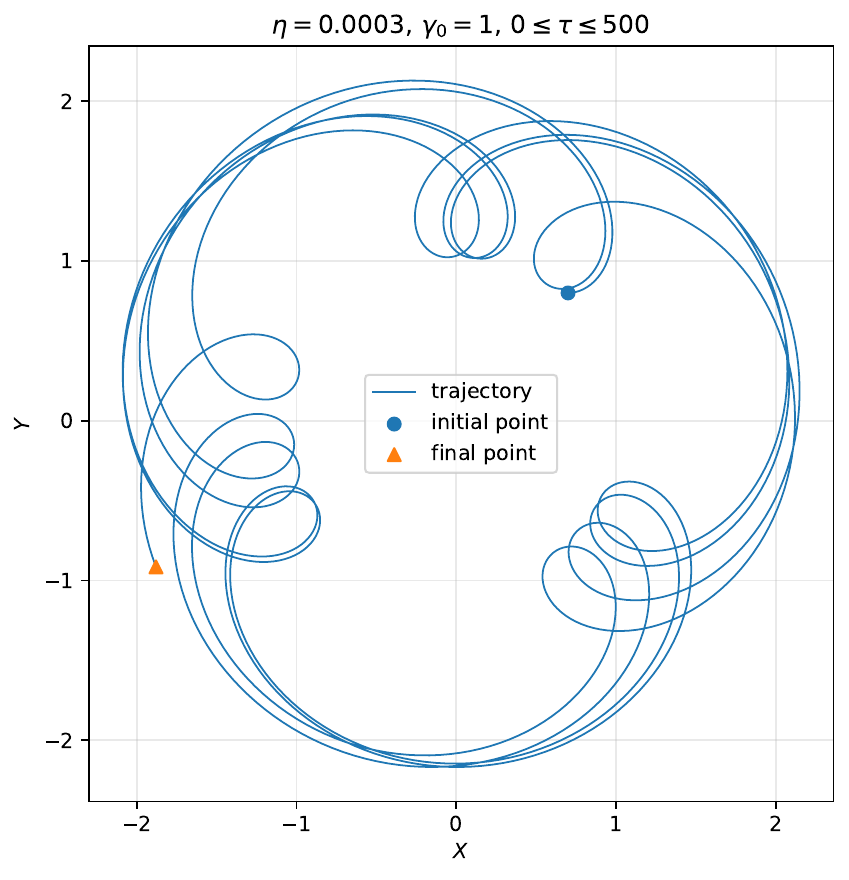}
	\hfill
	\includegraphics[width=0.48\textwidth]
	{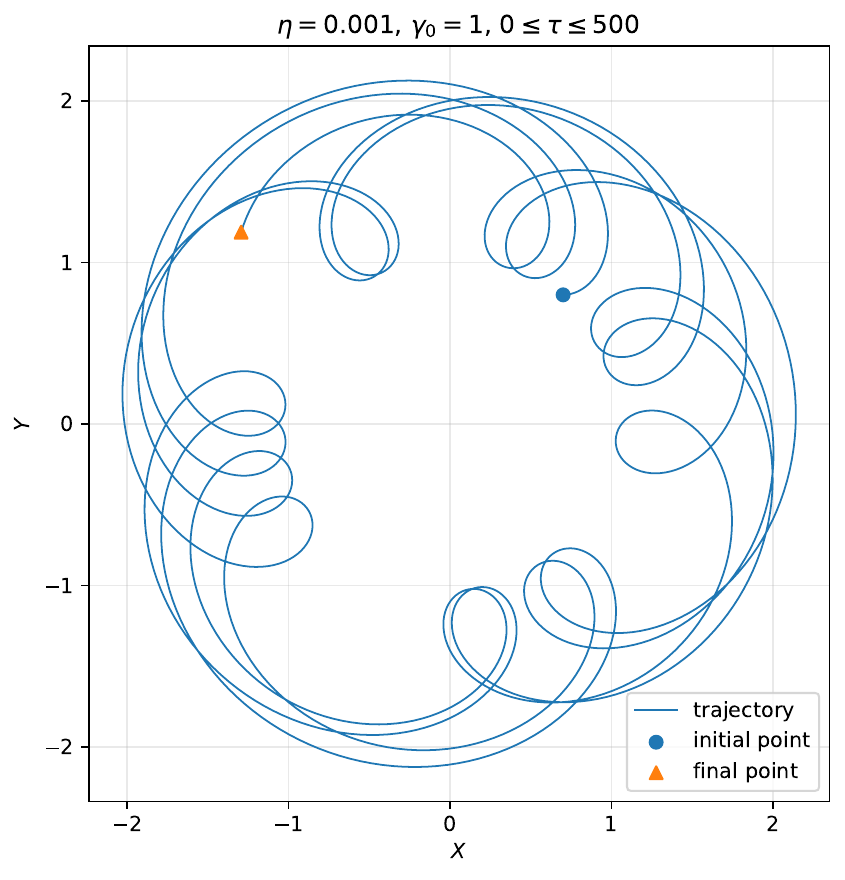}
	
	\includegraphics[width=0.48\textwidth]
	{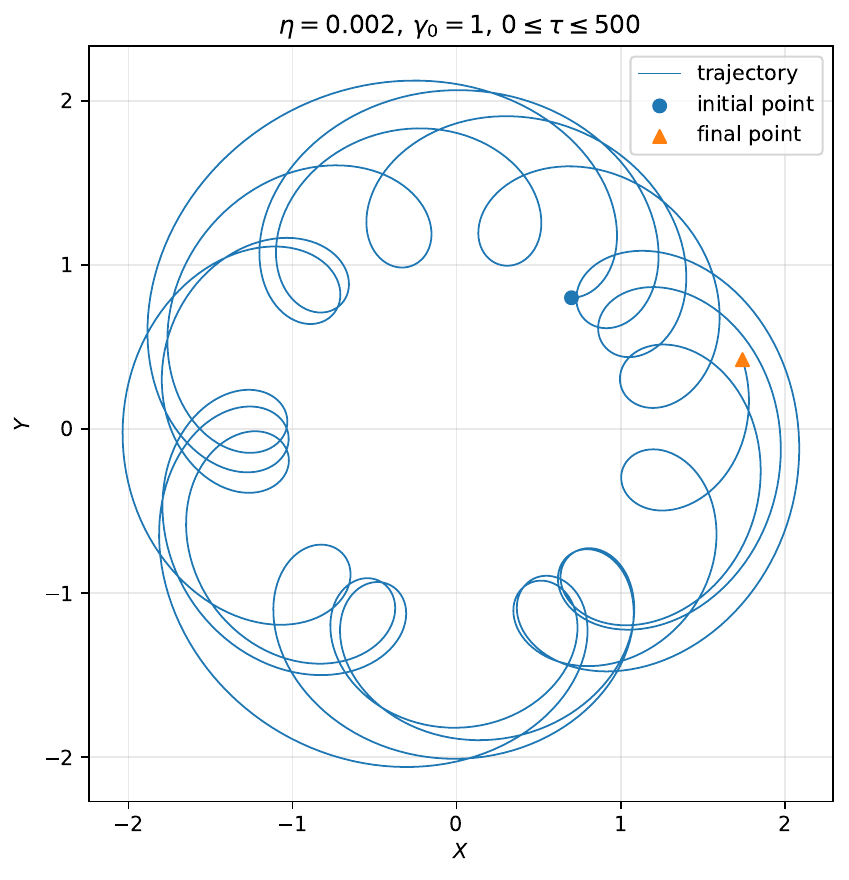}
	\hfill
	\includegraphics[width=0.48\textwidth]
	{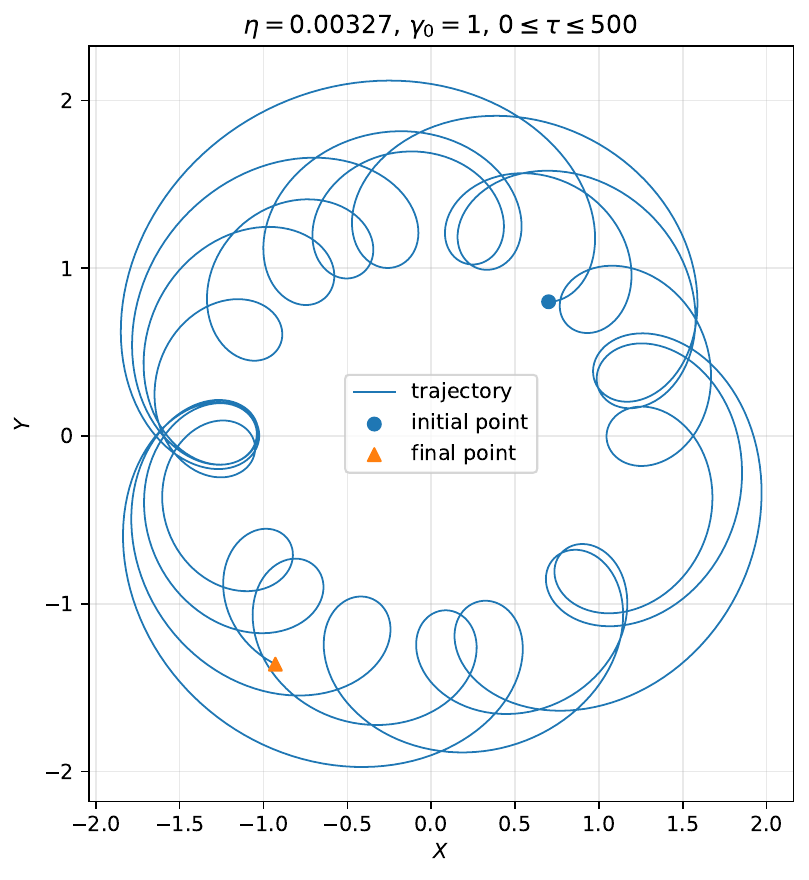}
	\caption{Representative equatorial trajectories of the reduced
		drag-only model for the initial data in
		Eq.~\eqref{eq:rr_num_initial_data}, with $\gamma_{0}=1$ and
		$0\leq\tau\leq500$. From upper left to lower right, the panels
		correspond to $\eta=3\times10^{-4}$, $10^{-3}$,
		$2\times10^{-3}$, and $3.27\times10^{-3}$. The circular and
		triangular markers indicate the initial and final positions,
		respectively. The differences between the paths reflect both the
		enhanced damping and the accumulated phase shift produced as $\eta$
		is increased.}
	\label{fig:rr_trajectories}
\end{figure*}

Figure~\ref{fig:rr_trajectories} shows integrations with
\begin{equation}
	\begin{aligned}
		X(0)
		&=
		0.7,
		&
		Y(0)
		&=
		0.8,
		\\[0.5ex]
		V_{X}(0)
		&=
		0.16,
		&
		V_{Y}(0)
		&=
		0.
	\end{aligned}
	\label{eq:rr_num_initial_data}
\end{equation}
We also set $Z(0)=V_{Z}(0)=0$ and $\gamma_{0}=1$. These data give
$\gamma(0)\simeq1.013051$ and
$p_{X}(0)\simeq0.162088$. The initial mechanical angular momentum is
negative because
$Xp_{Y}-Yp_{X}=-Y(0)p_{X}(0)<0$. The orbit therefore initially rotates in
the direction opposite to the positive circular branch discussed in
Sec.~\ref{sec:planar_decay}, and
Eq.~\eqref{eq:rr_num_K_decay} predicts an initial increase of
$\mathcal{K}$.

The integrations cover $0\leq\tau\leq500$ and use
$\eta=3\times10^{-4}$, $10^{-3}$, $2\times10^{-3}$, and
$3.27\times10^{-3}$. These values are deliberately much larger than the
illustrative astrophysical estimate in Eq.~\eqref{eq:eta}. Their purpose is
to make the cumulative dissipative deformation visible over a manageable
integration interval, rather than to represent a particular astrophysical
source.

The resulting paths are evolving rosette-like trajectories rather than
logarithmic spirals. The Lorentz force continually changes the momentum
direction, while radiation reaction changes its magnitude most efficiently
during close approaches to the dipole. The competition between these effects
produces a phase-dependent sequence of radial excursions. Increasing $\eta$
accelerates the energy loss and modifies the orbital phase at which each
inner passage occurs, so trajectories with different values of $\eta$ are
not related by a simple rescaling of time.

These examples illustrate finite-time dissipative transport in the reduced
planar model. They confirm that the motion can be substantially deformed
without following the formal circular-branch expansion derived in
Sec.~\ref{sec:planar_decay}. They do not establish a universal tendency
towards inward migration, outward migration, or circularisation.

\begin{figure}[t]
	\centering
	\includegraphics[width=\columnwidth]
	{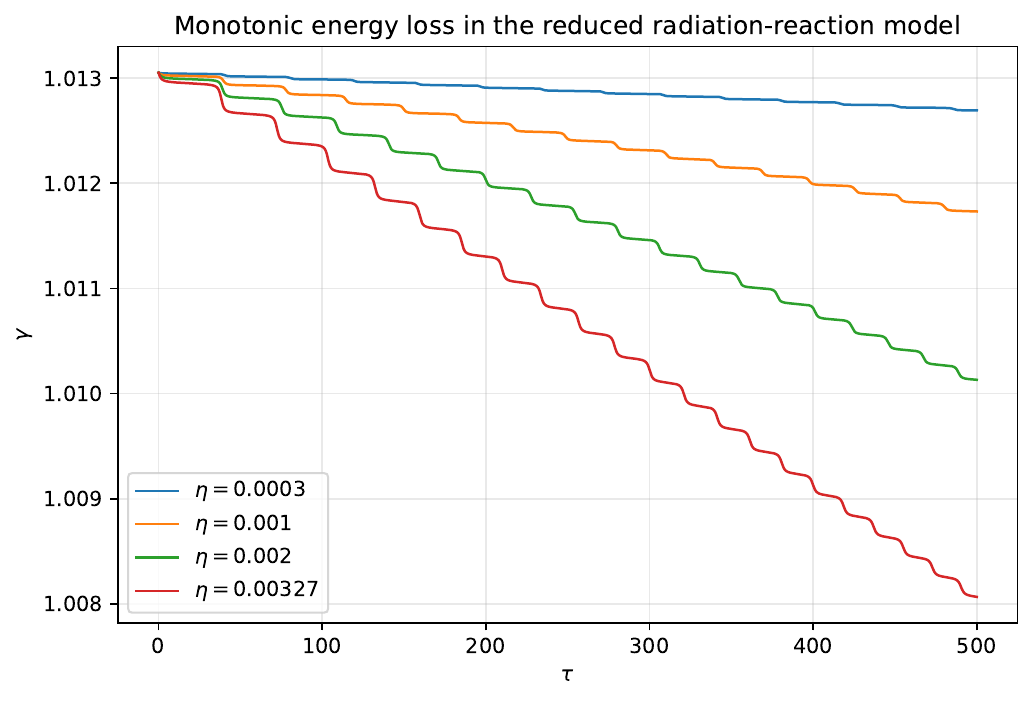}
	\caption{Evolution of the Lorentz factor for the four values of
		$\eta$ used in Fig.~\ref{fig:rr_trajectories}. In every case,
		$\gamma$ decreases monotonically, consistently with
		Eq.~\eqref{eq:rr_num_energy}. The nonuniform slopes arise from the
		strong radial dependence of the dipole-field damping rate.}
	\label{fig:rr_gamma_decay}
\end{figure}

Figure~\ref{fig:rr_gamma_decay} shows the corresponding Lorentz-factor
evolution. Its monotonic decrease provides a direct numerical check of
Eq.~\eqref{eq:rr_num_energy}. The variation of the slope reflects the
changing value of $R^{-6}$ along each orbit: rapid decreases occur during
inner passages, whereas the energy evolves more slowly when the particle
moves through the weaker outer field.

\section{Three-dimensional dynamics}
\label{sec:3D}

Outside the equatorial plane, the conservative CRSP admits regular and
chaotic trajectories with nontrivial vertical motion
\cite{Harko:2026tev}. Here we restrict the analysis to small transverse
perturbations of the instantaneous circular branch. The result is local: it
characterises the behaviour near that branch but does not establish that the
equatorial plane is a global attractor for arbitrary three-dimensional
trajectories.

\subsection{Linear transverse stability of the circular branch}
\label{sec:linear_stability}

Consider an instantaneous circular state of radius $R_{c}$,
\begin{equation}
	\begin{aligned}
		X_{c}
		&=
		R_{c}\cos\phi,
		&
		Y_{c}
		&=
		R_{c}\sin\phi,
		\\[0.5ex]
		Z_{c}
		&=
		0,
		&
		\dot{\phi}
		&=
		\omega
		=
		\frac{1}{\gamma R_{c}^{3}}.
	\end{aligned}
	\label{eq:circular_reference_3D}
\end{equation}
The azimuthal speed is
$V_{\phi}=\omega R_{c}=1/(\gamma R_{c}^{2})$. Under radiation reaction,
$R_{c}$ and $\gamma$ evolve on the slow dissipative timescale. They may
therefore be treated as locally constant during one vertical oscillation,
although their slow variation must be retained whenever it contributes at
order $\eta$.

The complete dimensionless momentum equation is
\begin{equation}
	\frac{\mathrm{d}\mathbf{p}}{\mathrm{d}\tau}
	=
	\mathbf{V}\times\mathbf{b}
	+
	\mathbf{A}_{\rm rad}^{\rm(full)},
	\label{eq:full_mom}
\end{equation}
where $\mathbf{A}_{\rm rad}^{\rm(full)}$ is given in
Eq.~\eqref{eq:Arad_full}. Let $Z=\zeta$, with
$|\zeta|\ll R_{c}$. Reflection symmetry about $Z=0$ implies that the
vertical perturbation decouples from the in-plane perturbations at linear
order. In cylindrical unit vectors,
\begin{equation}
	\begin{aligned}
		\mathbf{b}
		&=
		\frac{3\zeta}{R_{c}^{4}}\mathbf{e}_{R}
		-
		\frac{1}{R_{c}^{3}}\mathbf{e}_{z}
		+
		\mathcal{O}(\zeta^{2}),
		\\[0.5ex]
		\mathbf{V}
		&=
		V_{\phi}\mathbf{e}_{\phi}
		+
		\dot{\zeta}\mathbf{e}_{z}.
	\end{aligned}
	\label{eq:vertical_expansions}
\end{equation}
To this order,
$\mathbf{V}\cdot\mathbf{b}=-\dot{\zeta}/R_{c}^{3}$ and
$b^{2}=R_{c}^{-6}$. The vertical component of the conservative Lorentz term
is
\begin{equation}
	\left(
	\mathbf{V}\times\mathbf{b}
	\right)_{z}
	=
	-
	\frac{3\zeta}{\gamma R_{c}^{6}}.
	\label{eq:vertical_lorentz_term}
\end{equation}
This restoring force arises because a small displacement from the symmetry
plane generates a radial magnetic-field component. The azimuthal motion
across that component then accelerates the particle back towards $Z=0$.

The leading transverse damping comes from the field-gradient term of the
complete LL equation. To linear order,
\begin{equation}
	\left(
	\mathbf{V}\cdot\boldsymbol{\nabla}_{R}
	\right)
	\mathbf{b}=
	\frac{3\dot{\zeta}}{R_{c}^{4}}\mathbf{e}_{R}
	+
	\frac{3V_{\phi}\zeta}{R_{c}^{5}}\mathbf{e}_{\phi}.
\end{equation}
The second contribution arises from the variation of the cylindrical basis
along the circular orbit, even though the dipole-field components are
axially symmetric. Consequently,
\begin{equation}
	\left\{
	\eta\gamma
	\mathbf{V}\times
	\left[
	\left(
	\mathbf{V}\cdot\boldsymbol{\nabla}_{R}
	\right)
	\mathbf{b}
	\right]
	\right\}_{z}
	= -
	\frac{3\eta}{R_{c}^{6}}\dot{\zeta}.
	\label{eq:vertical_gradient_term}
\end{equation}

The first quadratic-field term has no linear vertical component,
$\left[
(\mathbf{V}\cdot\mathbf{b})\mathbf{b}
-b^{2}\mathbf{V}
\right]_{z}=0$, whereas the second contributes
$-\eta\gamma_{0}R_{c}^{-10}\dot{\zeta}$.

Since $p_{z}=\gamma\dot{\zeta}$, the linear vertical momentum equation is
\begin{equation}
	\frac{\mathrm{d}}{\mathrm{d}\tau}
	\left(
	\gamma\dot{\zeta}
	\right)
	=
	-
	\frac{3\zeta}{\gamma R_{c}^{6}}
	-
	\frac{3\eta}{R_{c}^{6}}\dot{\zeta}
	-
	\frac{\eta\gamma_{0}}{R_{c}^{10}}\dot{\zeta}.
	\label{eq:vertical_momentum_linear}
\end{equation}
Along the slowly evolving circular background,
Eq.~\eqref{eq:dg_circ} gives
$\mathrm{d}\gamma/\mathrm{d}\tau=-\eta\gamma_{0}/R_{c}^{10}$.
The corresponding contribution on the left-hand side cancels the final term
on the right-hand side of Eq.~\eqref{eq:vertical_momentum_linear}. The
vertical perturbation therefore satisfies
\begin{equation}
	\ddot{\zeta}
	+
	\frac{3\eta}{\gamma R_{c}^{6}}\dot{\zeta}
	+
	\frac{3}{\gamma^{2}R_{c}^{6}}\zeta
	=
	0.
	\label{eq:damped_vert}
\end{equation}

Equation~\eqref{eq:damped_vert} describes a damped harmonic oscillator. Its
conservative vertical frequency and amplitude-damping rate are
\begin{align}
	\Omega_{z}
	&=
	\frac{\sqrt{3}}{\gamma R_{c}^{3}}
	=
	\sqrt{3}\omega,
	\\
	\Gamma_{\perp}
	&=
	\frac{3\eta}{2\gamma R_{c}^{6}}.
	\label{eq:Gamma}
\end{align}
The vertical frequency is therefore of the same order as the orbital
frequency, whereas the damping is slower by a factor proportional to
$\eta$. This separation of timescales permits many vertical oscillations
during one damping time when radiation reaction is weak.

For locally constant coefficients and
$\Gamma_{\perp}<\Omega_{z}$, the underdamped solution is
\begin{equation}
	\begin{aligned}
		\zeta(\tau)
		&=
		A_{i}
		\exp\left[
		-\Gamma_{\perp}(\tau-\tau_{i})
		\right]
		\\
		&\quad\times
		\cos\left[
		\Omega_{d}(\tau-\tau_{i})+\varphi_{i}
		\right],
		\\[0.5ex]
		\Omega_{d}^{2}
		&=
		\Omega_{z}^{2}-\Gamma_{\perp}^{2}.
	\end{aligned}
	\label{eq:vert_sol}
\end{equation}
When $R_{c}$ and $\gamma$ vary slowly, the local envelope is proportional to
$\exp[-\int\Gamma_{\perp}(\tau)\mathrm{d}\tau]$.

The field-gradient force performs no work on the full three-dimensional
motion, but it may redistribute energy among the vertical and in-plane
degrees of freedom. Its damping effect on the vertical subsystem is therefore
consistent with the exact total energy balance, which is controlled by the
quadratic-field terms. This also explains why the reduced drag-only model
reproduces the total LL energy loss but misses the leading linear transverse
damping.

The local amplitude-damping time is
$\tau_{\rm damp}=2\gamma R_{c}^{6}/(3\eta)$, whereas the characteristic
timescale of the conditional circular-branch expansion is
$\tau_{\rm exp}=2R_{c}^{6}/(\eta\gamma)$. Their ratio is
\begin{equation}
	\frac{\tau_{\rm damp}}{\tau_{\rm exp}}
	=
	\frac{\gamma^{2}}{3}.
	\label{eq:damping_expansion_ratio}
\end{equation}
Transverse damping is therefore faster than the formal branch migration for
$\gamma<\sqrt{3}$, comparable near $\gamma=\sqrt{3}$, and slower for larger
Lorentz factors. This comparison concerns two local timescales along the
conditional circular branch and does not overcome its radial instability.

The positive coefficient $\Gamma_{\perp}$ establishes local transverse
damping under the complete LL equation. It does not prove that the
equatorial plane is a global attractor for arbitrary three-dimensional
trajectories. Away from the neighbourhood of the circular branch, radial
instability, nonlinear coupling, resonances, and chaotic transport may
compete with or modify the local planarising tendency.

\subsection{Local action-angle description of vertical damping}
\label{sec:nonlinear_vertical}

The linear analysis of Sec.~\ref{sec:linear_stability} establishes local
transverse damping near the circular branch when the complete
Landau--Lifshitz field-gradient term is retained. A one-dimensional
action-angle description may be introduced locally by treating the slowly
varying quantities $R_{c}$ and $\gamma$ as fixed during one vertical
oscillation and by neglecting coupling to the in-plane perturbations. This
construction describes the vertical dynamics in a neighbourhood of the
instantaneous circular state and is not an exact reduction of the full
three-dimensional system at finite amplitude.

With the canonical momentum $p_{Z}=\gamma\dot{Z}$, the frozen-background
Hamiltonian is
\begin{equation}
	H_{\perp}
	=
	\frac{p_{Z}^{2}}{2\gamma}
	+
	\gamma\Phi(Z;\gamma,R_{c}).
	\label{eq:H_perp}
\end{equation}
Within the approximation in which the cylindrical radius and azimuthal speed
are frozen at their circular values, the conservative vertical force can be
integrated to give
\begin{align}
	\Phi(Z;\gamma,R_{c})
	&=
	\frac{1}{\gamma^{2}R_{c}^{4}}
	\left[
	1-
	\left(
	1+\frac{Z^{2}}{R_{c}^{2}}
	\right)^{-3/2}
	\right]
	\nonumber\\
	&=
	\frac{1}{2}\Omega_{z}^{2}Z^{2}
	-
	\frac{5\Omega_{z}^{2}}{8R_{c}^{2}}Z^{4}
	+
	\mathcal{O}(Z^{6}),
	\label{eq:Phi_Z}
\end{align}
where
$\Omega_{z}=\sqrt{3}/(\gamma R_{c}^{3})$.
The quadratic term reproduces the frequency obtained in
Sec.~\ref{sec:linear_stability}. The negative quartic correction shows that
the local vertical oscillator is softening: its frequency decreases as the
oscillation amplitude grows.

The quartic truncation should not be extrapolated to large $|Z|$. In
particular, its apparent unboundedness from below is an artefact of truncating
the expansion. The unexpanded frozen-background potential remains bounded,
and the full three-dimensional dynamics also modifies the radial and
azimuthal motion at order $Z^{2}$. Equation~\eqref{eq:Phi_Z} is therefore
reliable only for sufficiently small transverse amplitudes.

For a periodic orbit of the frozen one-dimensional system, the canonical
vertical action is
\begin{equation}
	J_{Z}
	\equiv
	\frac{1}{2\pi}
	\oint p_{Z}\mathrm{d}Z.
	\label{eq:Jz_def}
\end{equation}
In the harmonic limit, $J_{Z}=H_{\perp}/\Omega_{z}$. At finite amplitude,
the action remains well defined as long as the frozen vertical motion is
periodic, although it must generally be evaluated by quadrature. The action
measures the phase-space area enclosed by one vertical oscillation and is
therefore a more natural diagnostic of transverse damping than the
instantaneous amplitude alone.

The vertical damping equation derived in Sec.~\ref{sec:linear_stability}
contains an order-$\eta$ cancellation between the second quadratic-field
term and the slow variation of the background Lorentz factor. After this
cancellation has been taken into account, and the remaining coefficients are
frozen over one vertical period, the leading dissipative contribution to the
vertical energy is
\begin{equation}
	\left.
	\frac{\mathrm{d}H_{\perp}}{\mathrm{d}\tau}
	\right|_{\rm diss}
	=
	-
	\frac{3\eta}{R_{c}^{6}}
	\dot{Z}^{2}
	+
	\mathcal{O}\left(
	\frac{\eta Z^{2}\dot{Z}^{2}}{R_{c}^{8}}
	\right).
	\label{eq:Hperp_decay}
\end{equation}
The leading term is negative for every nontrivial vertical oscillation. It
originates from the field-gradient part of the complete LL force and is
absent from the reduced drag-only model at linear order. Although the
field-gradient force performs no work on the full three-dimensional motion,
it transfers energy out of the vertical degree of freedom while the total
energy loss remains governed by the quadratic-field terms.

Let $\Omega_{\perp}(H_{\perp})$ denote the nonlinear angular frequency of the
frozen vertical orbit. Using
$\partial J_{Z}/\partial H_{\perp}=1/\Omega_{\perp}$ and averaging over one
vertical period gives
\begin{align}
	\left.
	\frac{\mathrm{d}J_{Z}}{\mathrm{d}T}
	\right|_{\rm diss}
	&=
	-\nu(H_{\perp})J_{Z},
	\label{eq:Jz_dot}
	\\
	\nu(H_{\perp})
	&=
	\frac{3}{R_{c}^{6}}
	\frac{
		\left\langle\dot{Z}^{2}\right\rangle
	}{
		\Omega_{\perp}J_{Z}
	},
	\label{eq:nu_nonlinear}
\end{align}
where $T=\eta\tau$, and the average is evaluated over the corresponding
frozen conservative orbit. Since
$\left\langle\dot{Z}^{2}\right\rangle>0$ for every nontrivial periodic
orbit, the dissipative contribution always decreases the vertical action
within this local model.

In the harmonic limit,
$\left\langle\dot{Z}^{2}\right\rangle=H_{\perp}/\gamma$ and
$\Omega_{z}J_{Z}=H_{\perp}$. Consequently,
\begin{equation}
	\nu(0)
	=
	\frac{3}{\gamma R_{c}^{6}}.
	\label{eq:nu_linear}
\end{equation}
In the original time variable, the action therefore decays at the rate
$\eta\nu(0)$, whereas the oscillation amplitude decays at half that rate,
$\Gamma_{\perp}=\eta\nu(0)/2$. This agrees with
Eq.~\eqref{eq:Gamma}, since the action is proportional to the square of the
amplitude in the harmonic regime.

The decay law in Eq.~\eqref{eq:Jz_dot} isolates the intrinsic dissipative
change of the vertical action at fixed $R_{c}$ and $\gamma$. In the complete
problem, the slow evolution of these background quantities produces
additional parametric changes, and nonlinear coupling to the radial and
azimuthal degrees of freedom may exchange action between the different
modes. Near resonances or separatrices, the single-mode averaging procedure
may also cease to be valid.

Thus the positivity of $\nu(H_{\perp})$ supports local planarisation near the
instantaneous circular branch, but it does not establish that the equatorial
plane is a global attractor. Large-amplitude and generic three-dimensional
trajectories must be studied by direct integration of the complete LL
equations, including the field-gradient term.

\subsection{Scope of the global three-dimensional evolution}
\label{sec:global_3D}

The analyses in Secs.~\ref{sec:linear_stability} and
\ref{sec:nonlinear_vertical} establish local transverse damping near the
instantaneous circular branch when the complete Landau--Lifshitz force is
retained. The local amplitude-damping time is
$\tau_{\rm damp}=2\gamma R_{c}^{6}/(3\eta)$, whereas the characteristic
timescale of the conditional circular-branch expansion is
$\tau_{\rm exp}=2R_{c}^{6}/(\eta\gamma)$. Their ratio is
\begin{equation}
	\frac{\tau_{\rm damp}}{\tau_{\rm exp}}
	=
	\frac{\gamma^{2}}{3}.
	\label{eq:ratio_global}
\end{equation}
Transverse damping is therefore faster than the formal branch migration for
$\gamma<\sqrt{3}$, comparable near $\gamma=\sqrt{3}$, and slower for
$\gamma>\sqrt{3}$. This comparison is local and assumes that $R_{c}$ and
$\gamma$ vary only slightly during one vertical oscillation.

The result does not imply that the equatorial plane is a global attractor.
The circular branch is radially unstable on the orbital timescale, whereas
both transverse damping and branch migration occur on the slower
radiation-reaction timescale. A trajectory may therefore depart radially
from the neighbourhood of the circular branch before the local vertical
damping has produced substantial planarisation. The relative ordering of
$\tau_{\rm damp}$ and $\tau_{\rm exp}$ does not address this radial
instability.

There is also an important distinction between the dynamical models used in
the planar and transverse analyses. The numerical planar integrations employ
the reduced drag-only equations, while the leading linear transverse damping
arises from the field-gradient term of the complete LL force. The reduced
model reproduces the exact instantaneous energy-loss law in a static magnetic
field, but it does not contain the complete directional dynamics required to
assess three-dimensional planarisation. Local vertical decay therefore
cannot be combined directly with the formal outward circular-branch solution
to predict the evolution of a generic three-dimensional orbit.

The planar results likewise establish neither generic circularisation nor a
finite terminal radius. The canonical radial action $J_{R}$ in
Eq.~\eqref{eq:J_R} is not generally conserved by the averaged dissipative
flow. Long-term planar transport is instead controlled by the coupled
evolution of $\gamma$ and $\mathcal{K}$ in
Eq.~\eqref{eq:slow_gammaK}, or by direct integration of the equations of
motion. The implicit solution in Eq.~\eqref{eq:tau_exact} remains an exact
analytical benchmark only for motion constrained to the instantaneous
circular branch.

The monotonic decrease of the orbital frequency and of the LL energy-loss
factor along that branch also does not determine the emitted radiation
spectrum. A spectral prediction requires the time-dependent radiated field
or an equivalent radiation calculation evaluated along the actual
trajectory. In particular, claims of a universal reverse chirp, a preferred
photon-energy range, or a specific ordering of early- and late-time spectral
power cannot be deduced from the orbital scalings alone.

For generic three-dimensional initial data, the dissipative drift may carry
the orbit through resonances, separatrices, and regions of chaotic motion
inherited from the conservative CRSP. Energy may also be exchanged among the
radial, azimuthal, and vertical degrees of freedom before being removed by
radiation reaction. Such mode coupling can temporarily enhance or oppose the
local transverse damping found near the circular branch.

A global assessment therefore requires numerical integration of the complete
LL equations, including the field-gradient term. Relevant diagnostics would
include the monotonic energy loss, the evolution of the canonical angular
momentum, radial and vertical actions where they remain well defined, the
distance from the equatorial plane, and the occurrence of resonant or
separatrix crossings. These quantities are needed to distinguish genuine
long-term planarisation from transient reductions of the vertical amplitude.

The firm conclusion of the present analysis is consequently local: small
vertical perturbations of the instantaneous circular branch are damped under
the complete LL dynamics. Global planarisation, generic circularisation, and
a universal late-time attractor have not been established.

\section{Discussion and conclusions}
\label{sec:discussion}

In this work, we have extended the classical relativistic St{\o}rmer problem
by incorporating radiation reaction within the Landau--Lifshitz
approximation. This provides a self-consistent framework in which the rapid
conservative motion generated by a magnetic dipole field is accompanied by a
much slower dissipative evolution. The formulation isolates the parameter
that controls the strength of radiation reaction and makes clear that its
physical magnitude depends sensitively on the magnetic dipole moment, the
particle species, and the length scale used to characterise the orbit.

An important outcome of the analysis is the distinction between the complete
Landau--Lifshitz dynamics and the reduced drag-only description. The
field-gradient contribution performs no work in a static magnetic field and
therefore does not affect the instantaneous energy-loss rate. It may,
nevertheless, alter the direction of the momentum and cannot be discarded in
general. The reduced model preserves the correct dissipative energy balance
and provides a useful approximation for nearly circular equatorial motion,
whereas the complete force is required whenever directional effects,
three-dimensional perturbations, or strongly nonuniform trajectories are
relevant.

For planar motion, radiation reaction causes the Lorentz factor and the
canonical angular momentum to evolve slowly across the family of conservative
orbits. Averaging over regular radial librations leads to a coupled secular
transport problem rather than to a single universal migration law. The
canonical radial action remains a valuable diagnostic of this transport, but
it is not generally conserved by the dissipative dynamics. Consequently, the
long-term behaviour of a librational orbit depends on the joint evolution of
its energy and angular momentum and cannot, in general, be reduced to a
one-parameter family of terminal states.

A particularly transparent analytical result is obtained when the motion is
constrained to remain on the instantaneous circular branch. Along this branch,
the loss of particle energy is accompanied by a gradual increase of the
orbital radius, and the resulting evolution can be integrated in closed form.
The asymptotic large-radius behaviour follows a simple power law and therefore
provides a useful benchmark for numerical calculations. This result must,
however, be interpreted conditionally: the conservative circular branch is
radially unstable, and its outward evolution does not represent the generic
response of arbitrary nearby trajectories. Nor does it lead to a finite
terminal radius; the weakening of the dissipation produces an increasingly
slow evolution that approaches the nonrelativistic regime only
asymptotically.

The planar numerical integrations complement this analytical picture. They
show how radiation reaction deforms rosette-like trajectories while producing
a monotonic decrease of the particle energy. Because the magnetic field varies
strongly along the orbit, the damping is highly nonuniform and cannot be
described by a constant exponential rate. The examples also demonstrate that
energy loss alone does not determine whether a general trajectory moves
inward or outward: the configuration-space evolution depends on the initial
conditions, the orbital phase, and the simultaneous drift of the conserved
quantities of the underlying conservative problem.

The three-dimensional analysis reveals a further role of the complete
Landau--Lifshitz force. Small vertical perturbations of the circular branch
are locally damped, and the leading damping mechanism originates precisely
from the field-gradient contribution that is absent from the reduced model.
A local action-angle description extends this conclusion to weakly nonlinear
vertical oscillations, provided that the background evolves slowly and the
coupling to the in-plane motion remains small. These results establish local
planarisation near the circular branch, but they do not imply that the
equatorial plane is a global attractor for arbitrary three-dimensional
trajectories. Radial instability, resonant coupling, separatrix crossing, and
chaotic motion may all compete with transverse damping outside the local
regime considered here.

The overall picture is therefore more structured than a universal
circularisation-and-escape scenario. The present study establishes a
hierarchy of controlled results: an exact relativistic energy-loss law, a
reduced planar model with a clearly identified domain of validity, averaged
transport equations for regular librations, a closed analytical solution
along the circular branch, and local transverse damping under the complete
Landau--Lifshitz force. Together, these results provide reliable analytical
benchmarks while identifying precisely where numerical investigation becomes
essential.

Several natural directions follow from this framework. A first priority is a
systematic comparison between trajectories generated by the complete and
reduced Landau--Lifshitz equations, with particular attention to the
field-gradient term, radial instability, and the onset of three-dimensional
motion. Global numerical surveys could map the fate of regular, resonant, and
chaotic initial conditions and determine whether long-lived transport
channels or statistically preferred regions of phase space emerge under weak
dissipation.

A second direction is the calculation of the electromagnetic radiation
produced by the trajectories themselves. The orbital-frequency and
energy-loss scalings derived here are not sufficient to determine an emitted
spectrum, but they provide the dynamical input required for such a
calculation. Combining the particle trajectories with the angular and
frequency distribution of the radiation would make it possible to identify
which dynamical features, if any, generate observable chirps, spectral
breaks, or transient signatures.

The single-particle analysis may also be extended to ensembles. Averaged
kinetic or Fokker--Planck descriptions could be constructed once the
phase-space drift induced by the complete self-force is sufficiently well
understood. Such models would permit the study of particle distributions,
radiative cooling, escape, and accumulation in inhomogeneous dipole fields,
and would provide a more direct connection with magnetospheric plasma
transport.

Finally, the present formulation provides a foundation for increasingly
realistic generalisations. Rotating electromagnetic fields, electric
components, curved spacetime, and frame-dragging may qualitatively modify the
balance between confinement and dissipation. In sufficiently strong fields,
quantum corrections to radiation reaction may also become important. The
classical relativistic St{\o}rmer problem developed here offers a controlled
starting point from which the influence of each of these effects can be
isolated and assessed.

In summary, radiation reaction enriches the relativistic St{\o}rmer problem
without eliminating the complexity inherited from its conservative dynamics.
Rather than producing a single universal late-time state, it drives a
phase-space transport process whose outcome depends on orbital geometry,
stability, and dimensionality. The combination of exact identities, local
analytical results, and reduced models obtained in this work provides a
coherent basis for future numerical, kinetic, and astrophysical studies of
charged-particle motion in strongly inhomogeneous magnetic fields.

\begin{acknowledgments}
FSNL acknowledges funding from the Funda\c{c}\~{a}o para a Ci\^{e}ncia e a Tecnologia (FCT) through national funds under the research grant UID/04434/2025 (DOI 10.54499/UID/04434/2025), and support from the FCT Scientific Employment Stimulus contract with reference CEECINST/00032/2018.
\end{acknowledgments}


\begin{thebibliography}{99}

\bibitem{St1} C.~St{\o}rmer, Arch. Sci. Phys. Nat. \textbf{24}, 5 (1907).
\bibitem{St2} C.~St{\o}rmer, Arch. Sci. Phys. Nat. \textbf{24}, 113 (1907).
\bibitem{St3} C.~St{\o}rmer, Arch. Sci. Phys. Nat. \textbf{24}, 221 (1907).
\bibitem{St4} C.~St{\o}rmer, Astrophys. J. \textbf{38}, 311 (1913).
\bibitem{St4a} C.~St{\o}rmer, Geofys.\ Publ.\ \textbf{1}, 269 (1921).
\bibitem{St5} C.~St{\o}rmer, Terr.\ Magn.\ Atmos.\ Electr.\ \textbf{22}, 97 (1917).
\bibitem{St6} C.~St{\o}rmer, Astrophys.\ Norv.\ \textbf{1}, 1 (1934).
\bibitem{St7} C.~St{\o}rmer, \emph{The Polar Aurora} (Clarendon Press, Oxford, 1955).

\bibitem{B1} A. Dragt, Rev. Geophys.  {\bf 3}, 255 (1965).

\bibitem{B2} A. Dragt and J. M. Finn, J. Geophys. Res.  {\bf 81}, 2327 (1976).

\bibitem{B3} M. Walt, \emph{Introduction to Geomagnetically Trapped Radiation}, Cambridge Atmospheric and Space Science Series, Cambridge University Press, 1994

\bibitem{Int} M. A. Almeida, I. C. Moreira, and H. Yoshida, J. Phys.
A Math. Gen. {\bf 25}, L227  (1992).

\bibitem{Dilao} R. Dil$\tilde{{\rm a}}$o and R. Alves-Pires , Chaos in the St\"{o}rmer Problem. In: Staicu, V. (eds) Differential Equations, Chaos and Variational Problems. Progress in Nonlinear Differential Equations and Their Applications, vol 75. Birkh\"{a}user, Basel, pp 175-194 (2007).

\bibitem{VA1} Y. Y. Shprits et al., Nature Physics \textbf{9}, 699 (2013).

\bibitem{VA2} J. H. Zhang, L. Y. Li, Y. W. Yao, K. X. Cheng, and L. Yang,
Journal of High Energy Astrophysics {\bf 52}, 100568 (2026).

 \bibitem{Schust}   R. Schuster and K. O. Thielheim, J. Phys. A: Math. Gen.  {\bf 20}, 5511 {\bf 1987}.
   
    \bibitem{How}  J. E. Howard, M. Hor\'{a}nyi, and G. R.  Stewart,   Phys. Rev. Lett.  {\bf 83}, 3993 (1999).
    
    \bibitem{Dull} H. R. Dullin,  M. Hor\'{a}nyi,  J. E.  Howard,  Physica D   {\bf 171}, 178 (2002).
    
     \bibitem{In} I$\tilde{{\rm n}}$arrea M. et al.,   Physica D {\bf  197}, 242 (2004).

\bibitem{In1} I$\tilde{{\rm n}}$arrea M. et al.,  Chaos, Solitons and Fractals  {\bf 42}, 155 (2009).

\bibitem{Epp0} V. Epp,  M. A.  Masterova,  Astrophysics and Space Science {\bf 353}, 473  (2014).

\bibitem{Epp} V. Epp,  O. N. Pervukhina,  Monthly Notices of the Royal Astronomical Society  {\bf 474}, 5330 (2018).

\bibitem{Hal} V. V. Markellos and A. A. Halioulias,  Astrophysics and Space Science {\bf  51},  177 (1977). 

\bibitem{Mark} V. V. Markellos and C. Zagouras,  Astronomy and Astrophysics  {\bf 61}, 505 (1977).

\bibitem{Ozturk} M. K. \"{O}zt\"{u}rk,  American Journal of Physics {\bf  80}, 420 (2012).


  \bibitem{Pina} E. Pina, E.  Cort\'{e}s,   European Journal of Physics {\bf  37},  065009 (2016).
  
  \bibitem{Kol} E. K. Kolesnikov,  Geomagnetism and Aeronomy {\bf 57},  137 (2017).
  
  \bibitem{Leg} A. Leghmouche, N. Mebarki,  A. Benslama,   New Astronomy {\bf 98},  101931 (2023).
  
  \bibitem{Ersh} S. Ershkov,  E. Prosviryakov, D. Leshchenko,  N. Burmasheva,   
 Mathematical Methods in the Applied Sciences {\bf  46},  19364 (2023).

\bibitem{Asadi} M. Asadi-Zeydabadi, C. S.  Zaidins,  Results in Physics  {\bf 12}, 2213 (2019). 


\bibitem{Ersh1} S. V. Ershkov,  J. Appl. Comput. Mech.  {\bf 12}, 31 (2026).

\bibitem{Moc} T. Harko and G. R. Mocanu, Annalen der Physik {\bf  537}, e00415 (2025).

\bibitem{Pap} D. B. Papadopoulos, I. Contopoulos, K. D. Kokkotas, N. Stergioulas, General Relativity and Gravitation {\bf 47}, 49 (2015). 

\bibitem{Bur} T. M. Burinskaya, M. M. Shevelev, Plasma Physics Reports {\bf 42}, 929 (2016). 

\bibitem{Bur1}  T. M. Burinskaya, M. M. Shevelev, Plasma Physics Reports {\bf 43}, 910 (2017).


\bibitem{Jack} J.~D.~Jackson, \emph{Classical Electrodynamics}, Wiley, Hoboken, NJ, 1999

\bibitem{LL} L.~D.~Landau and E.~M.~Lifshitz, \emph{The Classical Theory of Fields}, Pergamon Press, Oxford, 1994


\bibitem{Harko:2026tev}
T.~Harko and F.~S.~N.~
, Exact solutions, trajectories and radiation patterns in the classical relativistic St{\"o}rmer problem,
[arXiv:2605.04790 [astro-ph.HE]].


\bibitem{Th} 
K.~S.~Thorne,
Astrophys.\ J.\ Suppl.\ Ser.\ \textbf{8}, 1 (1963).



\bibitem{DiPiazza:2011tq}
A.~Di Piazza, C.~Muller, K.~Z.~Hatsagortsyan and C.~H.~Keitel,
Rev. Mod. Phys. \textbf{84},  1177 (2012).
[arXiv:1111.3886 [hep-ph]].

\bibitem{Barkov:2025uag}
M.~V.~Barkov and M.~Lyutikov,
[arXiv:2506.20515 [astro-ph.HE]].


\bibitem{Jerome:2022emr}
P.~J{\'e}r{\^o}me,
Astron. Astrophys. \textbf{666}, A5 (2022).
[arXiv:2207.00624 [astro-ph.HE]].


\bibitem{Tomczak:2023ftp}
I.~Tomczak and J.~P{\'e}tri,
Astron. Astrophys. \textbf{676}, A128 (2023).
[arXiv:2306.11482 [astro-ph.HE]].



\bibitem{Stuchlik:2024tlu}
Z.~Stuchl{\'\i}k, J.~Vrba, M.~Kolo{\v{s}} and A.~Tursunov,
JHEAp \textbf{44}, 500  (2024).
[arXiv:2412.04996 [astro-ph.HE]].


\bibitem{Cerutti:2015hvk}
B.~Cerutti, A.~A.~Philippov and A.~Spitkovsky,
Mon. Not. Roy. Astron. Soc. \textbf{457} 2401 (2016).
[arXiv:1511.01785 [astro-ph.HE]].

\bibitem{Philippov:2013tpa}
A.~A.~Philippov and A.~Spitkovsky,
Astrophys. J. Lett. \textbf{785}, L33  (2014).
[arXiv:1312.4970 [astro-ph.HE]].

\end{thebibliography}
\end{document}